\documentclass{aa} 

\usepackage[colorlinks = true,
            linkcolor = blue,
            urlcolor  = blue,
            citecolor = blue]{hyperref}
\usepackage[skip=0.5ex]{subcaption}
\usepackage{graphicx}
\usepackage{txfonts}
\begin{document}

    \title{Stellar mass-metallicity relation throughout the large-scale structure of the Universe: CAVITY mother sample
    }

    \subtitle{}
    
\author{Jes\'us Dom\'inguez-G\'omez\inst{1}
\and 
Isabel P\'erez\inst{1,2}
\and
Tom\'as Ruiz-Lara\inst{1,3}
\and
Reynier F. Peletier\inst{1,3}
\and
Patricia S\'anchez-Bl\'azquez\inst{4}
\and
Ute Lisenfeld\inst{1,2}
\and
Bahar Bidaran\inst{1}
\and
Jes\'us Falc\'on-Barroso\inst{5,6}
\and
Manuel Alc\'azar-Laynez\inst{1}
\and
Mar\'ia Argudo-Fern\'andez\inst{1,2}
\and 
Guillermo Bl\'azquez-Calero\inst{7}
\and
H\'el\`ene Courtois\inst{8}
\and
Salvador Duarte Puertas\inst{1,2,9}
\and
Daniel Espada\inst{1,2}
\and
Estrella Florido\inst{1,2}
\and
Rub\'en Garc\'ia-Benito\inst{7}
\and
Andoni Jim\'enez\inst{1}
\and
Kathryn Kreckel\inst{10}
\and
M\'onica Rela\~no\inst{1,2}
\and
Laura S\'anchez-Menguiano\inst{1,2}
\and
Thijs van der Hulst\inst{3}
\and
Rien van de Weygaert\inst{3}
\and
Simon Verley\inst{1,2}
\and
Almudena Zurita\inst{1,2}
}
\institute{Universidad de Granada, Departamento de F\'isica Te\'orica y del Cosmos, Campus Fuente Nueva, Edificio Mecenas, E-18071, Granada, Spain. \email{jesusdg@ugr.es}
\and
Instituto Carlos I de F\'isica Te\'orica y Computacional, Facultad de Ciencias, E-18071 Granada, Spain
\and
Kapteyn Astronomical Institute, University of Groningen, Landleven 12, 9747 AD Groningen, The Netherlands.
\and
Departamento de F\'isica de la Tierra y Astrof\'isica \& IPARCOS, Universidad Complutense de Madrid, E-28040, Madrid, Spain.
\and
Instituto de Astrof\'isica de Canarias, V\'ia L\'actea s/n, 38205 La Laguna, Tenerife, Spain.
\and
Departamento de Astrof\'isica, Universidad de La Laguna, 38200 La Laguna, Tenerife, Spain.
\and
Instituto de Astrof\'isica de Andaluc\'ia - CSIC, Glorieta de la Astronomía s.n., 18008 Granada, Spain
\and
Universit\'e Claude Bernard Lyon 1, IUF, IP2I Lyon, 4 rue Enrico Fermi, Villeurbanne, 69622, France
\and
D\'epartement de Physique, de G\'enie Physique et d’Optique, Universit\'e Laval, and Centre de Recherche en Astrophysique du Qu\'ebec (CRAQ), Qu\'ebec, QC, G1V 0A6, Canada
\and
Astronomisches Rechen-Institut, Zentrum f\"ur Astronomie der Universit\"at Heidelberg, M\"onchhofstraße 12-14, D-69120 Heidelberg, Germany.}

    \date{Accepted October 13, 2023}

 
    \abstract
    {Void galaxies are essential for understanding the physical processes that drive galaxy evolution because they are less affected by external factors than galaxies in denser environments, that is, in filaments, walls, and clusters. The stellar metallicity of a galaxy traces the accumulated fossil record of the star formation through the entire life of the galaxy. A comparison of the stellar metallicity of galaxies in various environments, including voids, filaments, walls, and clusters can provide valuable insights into how the large-scale environment affects the chemical evolution of the galaxy. 
    }
    {We present the first comparison of the relation of the total stellar mass versus central stellar metallicity between galaxies in voids, filaments, walls, and clusters with different star formation history (SFH) types, morphologies, and colours for stellar masses between $10^{8.0}$ to $10^{11.5}$ solar masses and redshift $0.01<z<0.05$. We aim to better understand how the large-scale structure affects galaxy evolution by studying the stellar mass-metallicity relation of thousands of galaxies, which allows us to make a statistically sound comparison between galaxies in voids, filaments, walls, and clusters. 
    }  
    {We applied non-parametric full spectral fitting techniques (pPXF and STECKMAP) to 10807 spectra from the SDSS-DR7 (987 in voids, 6463 in filaments and walls, and 3357 in clusters) and derived their central mass-weighted average stellar metallicity ($\rm [M/H]_M$). 
    }
    {We find that galaxies in voids have slightly lower stellar metallicities on average than galaxies in filaments and walls (by~$\sim~0.1$~dex), and they are much lower than those of galaxies in clusters (by~$\sim~0.4$~dex). These differences are more significant for low-mass ($ \sim~10^{9.25}~{\rm M_\odot}$) than for high-mass galaxies, for long-timescale SFH (extended along time) galaxies than for short-timescale SFHs (concentrated at early times) galaxies, for spiral than for elliptical galaxies, and for blue than for red galaxies.
    } 

   
    \maketitle
%
\section{Introduction}

The distribution of galaxies in the Universe forms a web-like structure with over-dense clusters, elongated filaments, sheetlike walls, and under-dense voids. This structure is detected by high-redshift galaxy surveys such as the Sloan Digital Sky Survey \citep[SDSS, ][]{2000AJ....120.1579Y}, the 2dF Galaxy Redshift Survey \citep[2dFGRS, ][]{2001MNRAS.328.1039C}, or the 2MASS Redshift Survey \citep[2MRS, ][]{2012ApJS..199...26H}. Voids occupy large volumes (from 10 to \makebox{$\rm30~h^{-1}~Mpc$} in diameter) with a low number density of galaxies \citep[density contrast \makebox{$\delta=\delta\rho/\rho\lesssim-0.8$}, where $\rho$ is the average density of the Universe, ][]{2001ApJ...557..495P, 2011AJ....141....4K, 2012AJ....144...16K, 2012MNRAS.421..926P, 2012ApJ...744...82V, 2016IAUS..308..493V}. Galaxies in voids are less affected by an intense gravitational potential and local processes than galaxies in filaments, walls, and clusters. These local processes include mergers or tidal interactions with other galaxies, as well as hydro-dynamical interactions between the intracluster medium (ICM) and the interstellar medium (ISM), such as ram-pressure stripping \citep{2006PASP..118..517B}. This makes voids a good probe to study the importance of internal processes in galaxy evolution.

Previous studies have shown that galaxies in voids are bluer, less massive, have later morphological types, and higher specific star formation rates (SFR) on average than galaxies in denser environments \citep{2004ApJ...617...50R, 2007ApJ...658..898P, 2012MNRAS.426.3041H, 2012AJ....144...16K, 2021ApJ...906...97F}. However, there is no consensus about SFR differences for a given stellar mass, luminosity, or morphology. Some studies have found that void galaxies have an enhanced SFR for a given stellar mass \citep{2005ApJ...624..571R, 2016MNRAS.458..394B, 2021ApJ...906...97F}, but others did not find any significant difference \citep{2006MNRAS.372.1710P,2012AJ....144...16K, 2014MNRAS.445.4045R, 2022A&A...658A.124D}. Additionally, \cite{2023MNRAS.524.5768P} have recently found that at low redshifts \makebox{($z<0.075$)}, the fraction of late-type galaxies is higher in voids than in the field, but these differences are not conclusive at higher redshifts \makebox{($0.075<z<0.150$).}

In our recent work \citep{2023Natur.619..269D}, we have compared the SFHs (as derived from a full-spectral fitting of the central parts) between galaxies located in different large-scale environments, and we found that galaxies in voids assemble their stellar mass more slowly than galaxies in filaments, walls, and clusters. Several physical processes might cause these differences: the different gas-accretion modes that dominate in each large-scale environment \citep{2005MNRAS.363....2K}, the lack of atomic hydrogen that might be present in void galaxies \citep{2012AJ....144...16K, 2022A&A...658A.124D, 2022MNRAS.tmp.2385R}, their higher halo-to-stellar mass ratio \citep{2018MNRAS.480.3978A, 2020A&A...638A..60A, 2020MNRAS.493..899H, 2022MNRAS.tmp.2385R}, their higher fraction of active galactic nuclei \citep[AGNs,][]{2021ApJ...922L..17M, 2022MNRAS.509.1805C}, their higher fraction of massive black holes \citep[BHs,][]{2008ApJ...673..715C}, and their lower local density \citep{2020A&A...639A..71K}.

There are numerous studies about the gas-phase mass-metallicity relation \makebox{($\rm MZ_g R$)} that only considered star-forming galaxies, but only a few studies focused on the stellar mass-metallicity relation \makebox{($\rm MZ_\star R$)} and also considered quenched galaxies. The gas metallicity is largely affected by the current star formation \citep{2014ApJ...797..126S, 2022A&A...666A.186D}, but the stellar metallicity traces the accumulated fossil record of the galaxy's star formation through its entire life. The comparison of the \makebox{$\rm MZ_\star R$} between galaxies in voids, filaments, walls, and clusters will help us to better understand how the large-scale environment affects the galaxy evolution.

Some studies have compared the \makebox{$\rm MZ_g R$} between different large-scale environments, but there is no consensus about the metallicity properties in void galaxies. \cite{2011AstBu..66..255P} found that the gas metallicity of dwarf void galaxies (absolute B-band magnitude, \makebox{$-18.4<M_{\rm B}<-11.9$)} is lower by about 30\% than in galaxies in denser large-scale environments, but \cite{2015ApJ...798L..15K} did not find any significant difference between dwarf galaxies in different large-scale environments. \cite{2008AJ....136....1W} found tentatively lower gas-phase metallicities in early-type void galaxies, but \cite{2019ApJ...883...29W} did not find any significant gas-phase metallicity difference between star-forming galaxies in voids and galaxies in denser large-scale environments. However, the robustness of these results is hampered by the low number of galaxies (20, 8, 26, and 33, respectively), which does not allow strong conclusions. Additionally, \cite{2008MNRAS.391.1117P} have shown that the gas-phase metallicity of cluster galaxies increases with their environmental density.

Other studies have analysed the effect of the local environment on the gas-phase chemical abundance of galaxies. \cite{2012MNRAS.425..273P} found that satellite galaxies have higher gas-phase metallicities than central galaxies with the same stellar mass, which is more significant at low than at high stellar masses, and the maximum differences lie (by \makebox{$\sim 0.06~{\rm dex}$)} at \makebox{$M_\star\sim 10^{8.25}~{\rm M_\odot}$}. Additionally, at a fixed stellar mass, the gas-phase metallicity of satellite galaxies increases with the halo mass \makebox{($M_{\rm h}$)} of the group, also more significantly for low-mass galaxies, and the maximum differences lie (by \makebox{$\sim 0.15~{\rm dex}$)} at \makebox{$M_\star\sim 10^{9}~{\rm M_\odot}$} inside the range \makebox{$10^{11}<M_h/{\rm M_\odot}<10^{14}$}. \cite{2011AJ....141..162D} found that star-forming galaxies with high \makebox{($-22.5\leq M_{\rm r}\leq -20.5$)} and low luminosities \makebox{($-20.5\leq M_{\rm r}\leq -18.5$)} have higher oxygen abundances in regions with higher local densities. \cite{2017MNRAS.465.1358P} found that late-type galaxies with higher local densities have higher oxygen (by \makebox{$\rm \sim 0.05~dex$)} and nitrogen (by \makebox{$\rm \sim 0.1~dex$)} abundances. This effect is more significant for low-mass than for high-mass galaxies. They derived the local density as the number of neighbours inside five different projected distances, $R_0=$ 1, 2, 3, 4, and 5~Mpc, but they did not compare galaxies in different large-scale structures such as voids, filaments, walls, and clusters. They also found that regions with the highest local ($R_0=$ 1~Mpc) densities are not necessarily associated with the highest large-scale ($R_0=$ 5~Mpc) densities, which supports the hypothesis that high local densities are also found in voids. This confirms that local and large-scale environments are not the same, and further comparisons are needed between galaxies in voids, filaments, walls, and clusters.

A few studies have analysed how the stellar metallicity of galaxies is affected by their local environment, but little is known about the effect of the large-scale environment. \cite{2021MNRAS.502.4457G} found that the stellar metallicity of cluster galaxies increases with halo mass, but they did not compare their results with void galaxies. \cite{2010MNRAS.407..937P} found that satellite galaxies have higher stellar metallicities than central galaxies with the same stellar mass. Additionally, the \makebox{$\rm MZ_\star R$} is shallower in systems with more massive haloes because the stellar metallicity of low-mass satellite galaxies increases with the halo mass of the system. These findings prove that the local environment of the galaxies affects the chemical evolution of their stars, but the effect of their large-scale environment remains unknown.

In this paper, we compare for the first time the \makebox{$\rm MZ_\star R$} between thousands of galaxies in voids, filaments, walls, and clusters. This study is linked to the Calar Alto void integral-field treasury survey (CAVITY\footnote{https://cavity.caha.es/}) project, which is an integral-field unit (IFU) legacy survey for void galaxies. It aims to observe around 300 galaxies with the PMAS-PPAK IFU of the Centro Astronómico Hispano en Andalucía (CAHA) together with ancillary deep-imaging, HI, and CO data {\color{blue}(Pérez et al. in
prep.)} to study the spatially resolved stellar populations, gas properties, and kinematics of void galaxies. As a complementary and preparatory study of this project, we derive here the average stellar metallicity of the void galaxy mother sample of CAVITY, and compare our results with galaxies in filaments, walls, and clusters. We apply non-parametric full spectral fitting techniques to the integrated optical spectra in the centre of the galaxies, which are already available in the SDSS, to obtain the stellar populations and metallicities.

This paper is organised in six sections and five appendices. In Section~\ref{sec:sample} we present the void galaxies and control samples of the study. In Section~\ref{sec:analysis} we describe the analysis we used to obtain the stellar metallicities. In Section~\ref{sec:results} we compare the \makebox{$\rm MZ_\star R$} between galaxies in different large-scale environments for different SFH types, morphologies, and colours. In Section~\ref{sec:discussion} we discuss our results and compare them with those of previous studies. In Section~\ref{sec:conclusions} we summarise our conclusions. In Appendix~\ref{sec:sample-KS} we extend our study to galaxy samples with the same stellar mass distribution. In Appendix~\ref{sec:flux-limited} we determine whether our results remain valid for volume-limited sub-samples with redshifts between 0.01 and 0.03. In Appendix~\ref{sec:s2n} we analyse whether there is any sample selection effect due to our cut in signal-to-noise ratio (S/N). In Appendix~\ref{sec:MZLR} we show similar results for the luminosity-weighted metallicities. In Appendix~\ref{sec:tables} we present the tables with our \makebox{$\rm MZ_\star R$} results. 

\section{Sample}\label{sec:sample}

We used the samples of galaxies defined in our previous work \citep{2023Natur.619..269D}, which were extracted from the spectroscopic catalogue of the SDSS-DR7 with redshifts \makebox{$0.01<z<0.05$.} We selected the mother sample of the CAVITY project as the mother sample of void galaxies in our study, with 2529 galaxies (see Section~\ref{sec:sample-cavity}). Our mother control sample is made of 6189 galaxies in clusters from \cite{2017A&A...602A.100T}, and 15000 galaxies in filaments and walls as they belong neither to voids nor to clusters (see Section~\ref{sec:sample-control}). We consider that filaments and walls belong to the same large-scale environment (filaments \& walls here after) as the number density of galaxies is very similar. After applying the spectral analysis (see Section~\ref{sec:analysis}) to these galaxies, we carried out a quality control (see Section~\ref{sec:QC}) and removed galaxies with low-quality spectra (\makebox{S/N$<20$}) from the mother samples, leaving us with 987 galaxies in voids, 6463 in filaments \& walls, and 3357 in clusters. Additionally, in Appendix~\ref{sec:sample-KS}, we define three sub-samples with the same total stellar mass distribution as our sample of void galaxies by applying the Kolmogorov-Smirnov test (KS-test). The total stellar masses of the galaxies were obtained from the database of the Max-Planck-Institut für Astrophysik and the Johns Hopkins University \citep[MPA-JHU,][]{2003MNRAS.341...33K, 2007ApJS..173..267S}. All these samples are magnitude limited by the SDSS completeness limit at \makebox{$r$-Petrosian $<$ 17.77 mag} \citep{2002AJ....124.1810S, 2015A&A...578A.110A}. 

    \subsection{Calar Alto void integral-field treasury survey} \label{sec:sample-cavity}
    
    The mother sample of the CAVITY comprises 2529 galaxies, which form a sub-sample of the \cite{2012MNRAS.421..926P} catalogue of SDSS void galaxies. \cite{2012MNRAS.421..926P} applied the VoidFinder algorithm \citep{1997ApJ...491..421E, 2002ApJ...566..641H} to the distribution of SDSS galaxies with redshifts \makebox{$z<0.107$}, and found 79947 void galaxies inside 1055 cosmic voids with a typical density contrast \makebox{$\delta=-0.94\pm0.03$} and radii larger than \makebox{$10~{\rm h^{-1}~Mpc}$.} The CAVITY collaboration {\color{blue}(Pérez et al. in prep.)} reduced the redshift range \makebox{($0.01<z<0.05$)} to concentrate on nearby galaxies that are observable with PMAS-PPAK. They chose 15 voids with more than 20 galaxies each to observe around 300 galaxies that are distributed along the entire right ascension range of the SDSS. They finally selected galaxies in the inner region of the voids (i.e. inside 80\% of the effective radius of the void) to avoid the possible inclusion of galaxies that inhabit or are affected by denser environments. Additionally, the CAVITY collaboration carried out a visual inspection of the galaxies, and removed duplicated objects from the sample and the spectra integrated in \makebox{H{\tiny II}} regions, not in the centre of the galaxy. 
    
    \subsection{Control samples} \label{sec:sample-control}
    
    The aim of this study is to compare the stellar metallicities between galaxies in voids and galaxies in denser environments. To do this, we defined two control samples: one sample of galaxies in clusters, and the other sample of galaxies in filaments \& walls. The mother sample of galaxies in clusters was extracted from the \cite{2017A&A...602A.100T} catalogue of groups of SDSS galaxies within the same redshift range as the CAVITY mother sample. Galaxies in groups with \makebox{$\geq 30$} members were selected as cluster galaxies \citep[][ see Appendix \ref{sec:flux-limited} for some discussions on how this criterium might affect our results]{1989ApJS...70....1A}. With these selection criteria, our mother sample of cluster galaxies contains 6189 galaxies.
    
    The mother sample of galaxies in filaments \& walls in this study was extracted from all the SDSS galaxies within the same redshift range as the mother sample of CAVITY that are neither in the complete catalogue of void galaxies of \cite{2012MNRAS.421..926P} nor in the mother sample of cluster galaxies defined above. To save computational time, we selected a sub-sample of 15000 galaxies in filaments \& walls, preserving similar stellar mass, $g-r$ colour, and redshift distributions (two-sample KS-test with \makebox{p-values > 0.95)} as the original sample of galaxies in filaments \& walls directly extracted from SDSS.

\section{Data analysis}\label{sec:analysis} 

In \cite{2023Natur.619..269D}, we carried out a non-parametric full spectral fitting analysis to derive the stellar populations and compare the SFHs of galaxies in voids, filaments \& walls, and clusters. Now, we compare the stellar metallicities of these three galaxy samples (see Section \ref{sec:sample}) using the same stellar populations. This spectral analysis recovers the stellar line-of-sight velocity distribution (LOSVD) and gas emission lines, and it generates a combination of stellar population models that best fit the observed spectra of the galaxies in a wavelength range from 3750 to 5450~\AA, in which the most relevant absorption lines of the stars are located. From this combination of models, we can estimate the mass, age, and metallicity of the stars within the galaxies.

For the analysed data, we used optical spectra from the SDSS-DR7 \citep{2009ApJS..182..543A} integrated (fibre aperture with a diameter of 3~arcsec) in the very centre of the galaxies (from 0.3 to 1.6~kpc in the redshift range of \makebox{$0.01<z<0.05$)} observed at the Apache Point Observatory (APO) 2.5~m telescope. We used the stellar models of the extended medium resolution INT library of empirical spectra \citep[E-MILES, ][]{2006MNRAS.371..703S, 2011A&A...532A..95F, 2015MNRAS.449.1177V, 2016MNRAS.463.3409V}, which are single-age and single-metallicity stellar population (SSP) spectral templates generated assuming the BaSTI  isochrones \citep{2004ApJ...612..168P} and Kroupa universal initial mass function \citep[IMF, ][]{2001MNRAS.322..231K}.
With the penalized pixel-fitting \citep[pPXF,][]{2004PASP..116..138C, 2017MNRAS.466..798C, Cappellari2022} algorithm, we generated a combination of stellar population models (E-MILES SSPs) and pure Gaussian emission line templates that best fit the observed spectra of the galaxies, recovering the stellar LOSVD and gas emission. The gas emission obtained by pPXF was subtracted from the observed spectrum of the galaxy to obtain a clean spectrum with only the emission from the stars. Afterwards, we applied the stellar content and kinematics from high-resolution galactic spectra via maximum a posteriori algorithm \citep[STECKMAP, ][]{2006MNRAS.365...46O, 2006MNRAS.365...74O} to recover the stellar populations (stellar mass, age, and metallicity) of a galaxy by fitting a combination of E-MILES SSPs (as for pPXF) to the clean spectrum of the galaxy (only emission from the stars), assuming a fixed stellar LOSVD (previously derived with pPXF). We estimated the errors of the stellar populations (stellar mass, age, and metallicity) as the standard deviation of five Monte Carlo iterations from STECKMAP. For each Monte Carlo solution, we used as input the observed spectrum plus a spectrum of noise with the same standard deviation in the continuum. With this method, we analysed different spectra with the same signal and the same level of noise, but a different distribution the noise.

The recovered stellar populations are affected by the age-metallicity degeneracy of the stars \citep{1994ApJS...95..107W}, and young metal-poor galaxies might have been classified as old metal-rich stars. This effect arises because old stars with low metallicities have similar spectra as young stars with high metallicities. \cite{2011MNRAS.415..709S} analysed the effect of the age-metallicity degeneracy for STECKMAP using synthetic spectra for ages 1 and 10~Gyr and solar metallicity (0~dex), and compared them with other spectral index techniques. Their Figure~7 shows that the age-metallicity degeneracy effect is much more reduced in the case of STECKMAP (ages of $\sim1.00\pm0.04~{\rm Gy}$ and $\sim11\pm1~{\rm Gy}$, respectively, and metallicities of $\sim0.02\pm0.04~{\rm dex}$) than in the case of spectral indices (ages of $\sim1.0\pm0.1~{\rm Gy}$ and $\sim10\pm5~{\rm Gy}$, respectively, and metallicities of $\sim-0.1\pm0.2~{\rm dex}$).

The stellar and gas migration may influence the central metallicities, especially if and when there is a bar in a galaxy. Previous studies found that the gas-phase metallicity \citep{2011MNRAS.416.2182E} and stellar metallicities \citep{2011A&A...529A..64P} in the centre of barred galaxies are higher than those of unbarred galaxies, but \cite{2014MNRAS.442.2496C} did not find significant difference in metallicity (neither gaseous nor stellar) between barred and unbarred galaxies. Nevertheless, there is no evidence whether the fraction of barred galaxies is different in voids compared to denser environments.

    \subsection{Quality control \label{sec:QC}}

    The quality of the outcome of spectral fitting techniques is affected by the S/N in the continuum ($\rm 6000-6100~\AA$, rest frame) and the intensity of the emission lines, among others. A good indicator of the quality of the spectral fit is the residual spectrum, which is the difference between the observed and fitted spectrum. If the residuals are high, the observed spectrum is noisy, or the fitted spectrum is not a perfect match to the observed one. We removed a fraction of galaxies (61\% in voids, 57\% in filaments \& walls, and 46\% in clusters) with \makebox{S/N$<20$} from our sample\footnote{We tested that the results presented in this work are not contingent upon this S/N cut choice; see Appendix \ref{sec:s2n} for more details.} for which the residuals are higher than 2\% of the continuum level around $\rm H\beta$. This selection by S/N removes mainly low-mass galaxies, but we do not expect that it introduces any bias in our sample because the mean stellar mass of the removed galaxies is similar in the three environments ($\rm 10^{9.2\pm0.1}~M_\odot$ in voids, $\rm 10^{9.3\pm0.1}~M_\odot$ in filaments \& walls, and $\rm 10^{9.5\pm0.1}~M_\odot$ in clusters). However, even after this cut, the stellar mass distribution of the three samples is different. To avoid any effect of the difference stellar mass distributions in our results, we compare in Section \ref{sec:results} the stellar metallicities for stellar mass bins of 0.5 dex width. However, the distributions may be still slightly different within each stellar mass bin. To verify that this does not introduce any effect in our analysis for a given stellar mass, we therefore define in Appendix \ref{sec:sample-KS} three sub-samples with the same stellar mass distribution within each stellar mass bin by applying the KS-test, and we analyse the stellar metallicities of these sub-samples. In addition, we also compare in Appendix \ref{sec:sample-KS} the stellar metallicities for narrower stellar mass bins of 0.25 dex width. We find similar results to those derived from the entire sample and bins of 0.5 dex width.
    
    Some galaxies have a high S/N \makebox{($>20$)} spectrum and residuals lower than 2\% of the continuum level, but higher than the level of noise. This is due to a poor fit of the gas-emission lines. The pPXF algorithm is not efficient in fitting wide or asymmetric emission lines and may leave wavy features in the clean spectrum that consequently affect the STECKMAP fit and lead to incorrect stellar populations. We remov  small fractions of galaxies (5\% in voids, 8\% in filaments \& walls, and 7\% in clusters) with a high S/N \makebox{($>20$)} spectrum and residuals twice higher than the level of noise over \makebox{$\rm H\beta$} after subtracting the emission lines from our samples. After a careful visual inspection, we confirm that these galaxies have intense, wide, and asymmetric emission lines that pPXF is not able to fit properly. These small percentages of removed galaxies do not introduce any bias in our analysis.
    
    The optical spectra from SDSS are integrated over the central region of the galaxies. This might introduce a bias for galaxy samples with a wide redshift range, where this aperture would cover a large fraction of remote galaxies, but only the inner part of nearby ones. However, the redshift range of our samples is rather narrow \makebox{($0.01-0.05$)}, and their apparent size \makebox{($r$-Petrosian} radius, \makebox{$\rm R_{90r}$}, from SDSS) distributions is very similar for the three environments a\citep[see ][and Extended Data Figure 6 therein]{2023Natur.619..269D}. In order to minimise a possible size effect in our study, we removed a small fraction (1\% in voids, 4\% in filaments \& walls, and 5\% in clusters) of galaxies from our samples with \makebox{$\rm R_{90r}>20~arcsec$}, for which the spectrum would be relatively more influenced by the fibre aperture. After this quality control, we are left with 987 galaxies in voids, 6463 in filaments \& walls, and 3357 in clusters. These are the samples we study in this work.

    \subsection{Averaged stellar metallicity\label{sec:ANA_met}}

    The stellar populations recovered by STECKMAP are characterised by their stellar mass, stellar age, and the metallicity of the gas from which the stars were formed, providing an estimate of the type of stars that currently form a galaxy. In the E-MILES models, the metallicity is defined as the fraction of metals ($\rm Z$) normalised to the solar value ($\rm Z_\odot=0.0198$) as \makebox{$\rm [M/H]=\log_{10}(Z/Z_\odot)$}. We can then derive the mass-weighted average stellar metallicity of the galaxy as
    \begin{equation}
        {\rm[M/H]_M}=\frac{\sum M_\star{\rm[M/H]}_\star}{\sum M_\star}, 
        \label{eq:MW-met}
    \end{equation}
    
    \noindent where $M_\star$ and $\rm[M/H]_\star$ are the mass and metallicity of the stellar populations that form the galaxy, respectively, which are obtained through spectral fit. The BaSTI theoretical isochrones that are used by the E-MILES models cover the metallicity range \makebox{$\rm-2.27\leq[M/H]_\star\leq0.4$}. We find that the stellar metallicity values of some galaxies saturate at the upper boundary of the stellar models. The central parts of the galaxies can be metal rich, \makebox{$\rm0.30<[M/H]_M<0.47$} \citep[][see Section B therein]{2015A&A...581A.103G}, with values higher than the stellar models, leading to a saturation effect. We derived the stellar metallicities of the centre (3~arcsec diameter) of nearby galaxies \makebox{($0.01<z<0.05$)} using the integrated spectrum in the innermost regions of the galaxies (aperture from 0.3 to 1.6~kpc). For some galaxies, the stars in these inner regions are very metal rich, and the average stellar metallicity of the galaxy reaches the limit of the stellar models. 
    
    This effect was also found by \cite{2005MNRAS.362...41G} (see Figure~8 therein). However, they obtained average metallicities that are lower than ours because they analysed a sample of galaxies with a wider redshift range \makebox{($0.005<z<0.22$)}, and the SDSS spectra that they used were integrated over more external regions of the galaxies, where the stars are more metal poor than in the centre. In our results, the fraction of galaxies with this saturation issue in each stellar mass bin, SFH type, and environment is lower than 10\%, except for filament \& wall and cluster galaxies with short-timescale SFHs (see Sectio~\ref{sec:ANAsfh}) at high stellar masses \makebox{($M_\star>10^{10.5}~{\rm M_\odot}$)}, for which the fraction of saturated galaxies is about 20\%. There is no saturation for void galaxies. The saturation effect is conservative because it reduces the average stellar metallicity of filament \& wall and cluster massive galaxies, but does not affect void galaxies. This dilutes the stellar metallicity differences that we find at high stellar masses (see Section~\ref{sec:results}).

    \subsection{Star formation histories\label{sec:ANAsfh}}
    
\begin{figure}
        \centering
        \includegraphics[width=0.5\textwidth]{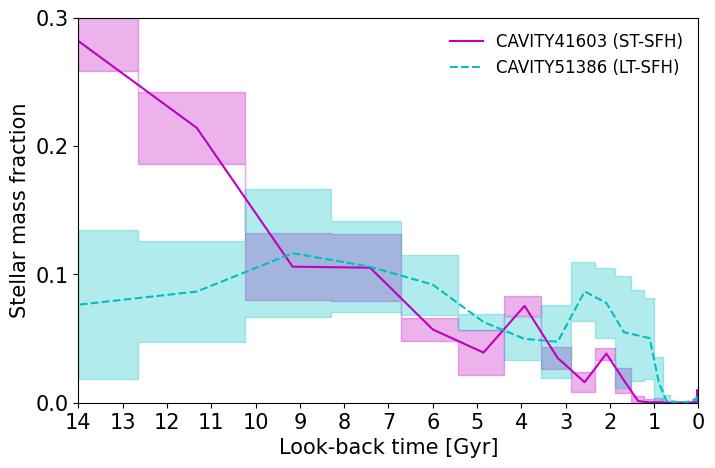}
        \caption{Examples of SFH as the stellar mass fraction vs. the look-back time of galaxies CAVITY41603 (solid magenta line) and CAVITY51386 (dashed cyan line), which are examples of ST-SFH and LT-SFH, respectively. The shaded boxes represent the uncertainties of the stellar mass fraction of the galaxy associate to each SSP that is representative for each look-back time bin. The lines represent the interpolation between the nominal values of the SSP mass fractions.
        }
        \label{fig:SFHexamples}
    \end{figure}

    The metallicity of a galaxy is strongly influenced by its SFH \citep{2002A&A...388..396T}. A high SFR quickly enriches the ISM and the stars that will form from it. Therefore, the effect of the SFH has to be considered in our analysis of the stellar metallicities. We derived and discussed the SFHs in our previous work \citep{2023Natur.619..269D} as the fraction of stellar mass that was formed at a given look-back time, which was based on the same spectral analysis. We found that the SFHs at early times describe a bimodal distribution in the three large-scale environments \citep[see][and Figure~2 therein for more details]{2023Natur.619..269D} , which  allowed us to classify the SFHs into two types: short-timescale SFH (ST-SFH) galaxies formed a large fraction of their stellar mass (27\% on average) \makebox{$\sim$~12.5~Gyr} ago and progressively reduced their star formation since then, while long-timescale SFH (LT-SFH) galaxies formed a lower fraction of their stellar mass (<~21.4\%) than the ST-SFH galaxies 12.5~Gyr ago, but formed stars more uniformly over time (see examples in Figure \ref{fig:SFHexamples}). 
    By definition, ST-SFH galaxies formed their stellar mass earlier than LT-SFH galaxies. Therefore, this classification needs to be taken into account when the SFHs between galaxies in different environments are compared.

    It is more likely for galaxies in voids to have an LT-SFH ($51.7\pm0.9\%$) than for galaxies in filaments \& walls ($44.5\pm0.3\%$) and clusters ($36.1\pm0.5\%$). For a given SFH type, galaxies in voids formed their stars more slowly on average than in filaments \& walls at intermediate stellar masses, and much more slowly than in clusters at any given stellar mass. The SFH differences between galaxies in the three large-scale environments might have affected the stellar metallicities diversely in voids, filaments \& walls, and clusters. Therefore, we analyse the effect of the large-scale environment on the stellar metallicities of the galaxies for different SFH types.

    \subsection{Morphology and colour\label{sec:ANAmorcol}}

    It might be intuitive to associate the ST-SFH type with red elliptical galaxies and the LT-SFH type with blue spiral galaxies. However, the SFH type of a galaxy clearly correlates neither with its colour nor its morphology \citep[see][and Extended Data Figure~7 therein for more details]{2023Natur.619..269D}. We therefore also analysed the \makebox{$\rm MZ_\star R$} for different morphologies and colours. We used the $g$ and $r$ dereddened magnitudes from SDSS to define the colour of the galaxies as $g-r$, and the T-type parameter from \cite{2018MNRAS.476.3661D} to characterise the morphology of the SDSS galaxies. We considered that galaxies with T-type~<~0 are elliptical and galaxies with T-type~>~0 are spiral \citep{2018MNRAS.476.3661D}, and we define galaxies with \makebox{$g-r\leq0.7~\rm{mag}$} as blue and \makebox{$g-r>0.7~\rm{mag}$} as red.  The SDSS $g$ and $r$ magnitudes are integrated over the entire galaxy, but the stellar populations are recovered from the very centre of the galaxy, where the stars are redder than the average colour $g-r$ of the entire galaxy.  
    
\section{Results}\label{sec:results}

\subsection{Distribution of the average stellar metallicity}

\begin{figure*}
    \centering
    \includegraphics[width=\textwidth]{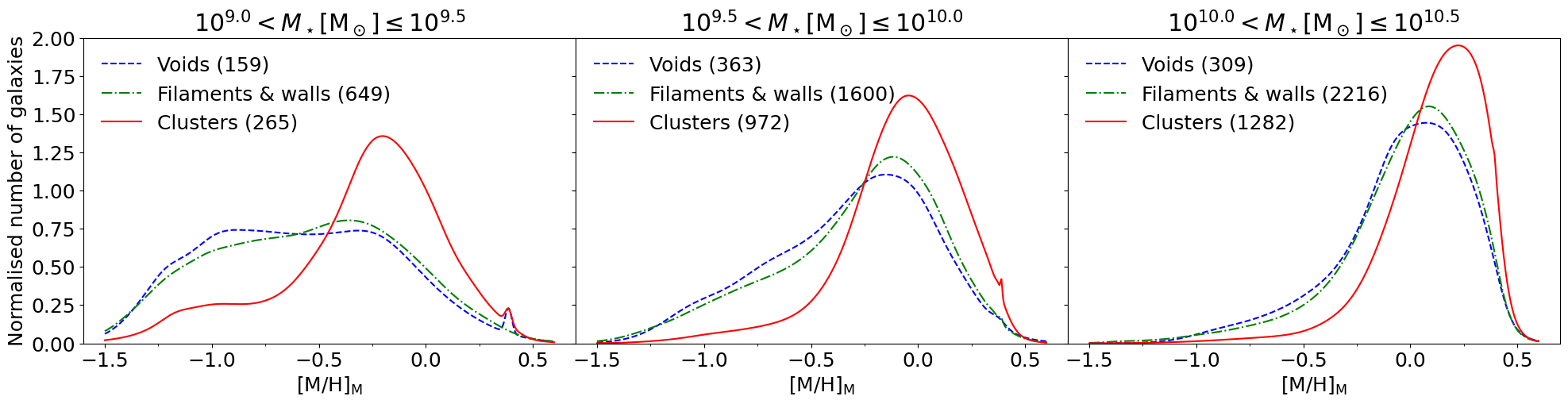}
    \caption{Mass-weighted average stellar metallicity ($\rm[M/H]_M$) distribution of our samples of galaxies in voids (dashed blue line), filaments \& walls (dot-dashed green line), and clusters (solid red line) for three stellar mass bins as labelled. The number of galaxies in each sample is shown in the legend. The peaks at $\sim0.4$ are due to the metallicity limit of the E-MILES stellar models. The distribution sharply ends at higher values ($\sim0.5$) than the limit due to the widening by the errors.}
    \label{fig:Z_dist}
\end{figure*}

We show in Figure~\ref{fig:Z_dist} the normalised distribution of $\rm[M/H]_M$ for galaxies in voids (dashed blue line), filaments \& walls (dot-dashed green line), and clusters (solid red line) in three total stellar mass bins. The small peaks at \makebox{$\rm[M/H]_M\sim0.4$,} are produced by the saturation of our results at the upper metallicity limit of the E-MILES SSPs. The distributions exhibit a sharp cut-off at higher values (around 0.5) due to the broadening effect caused by the errors. We find that the stellar metallicity distribution is similar for galaxies in voids and galaxies in filaments \& walls. However, cluster galaxies are distributed at much higher metallicities. These differences might be affected by their different stellar mass distributions and SFH types. Void galaxies are less massive on average and form their stars more slowly than galaxies in denser environments \citep{2023Natur.619..269D}. We then compared different stellar mass bins, SFH types, morphologies, and colours.

\subsection{Stellar mass effect}\label{sec:mstar_eff}

\begin{figure*}

    \centering
    \includegraphics[width=\linewidth]{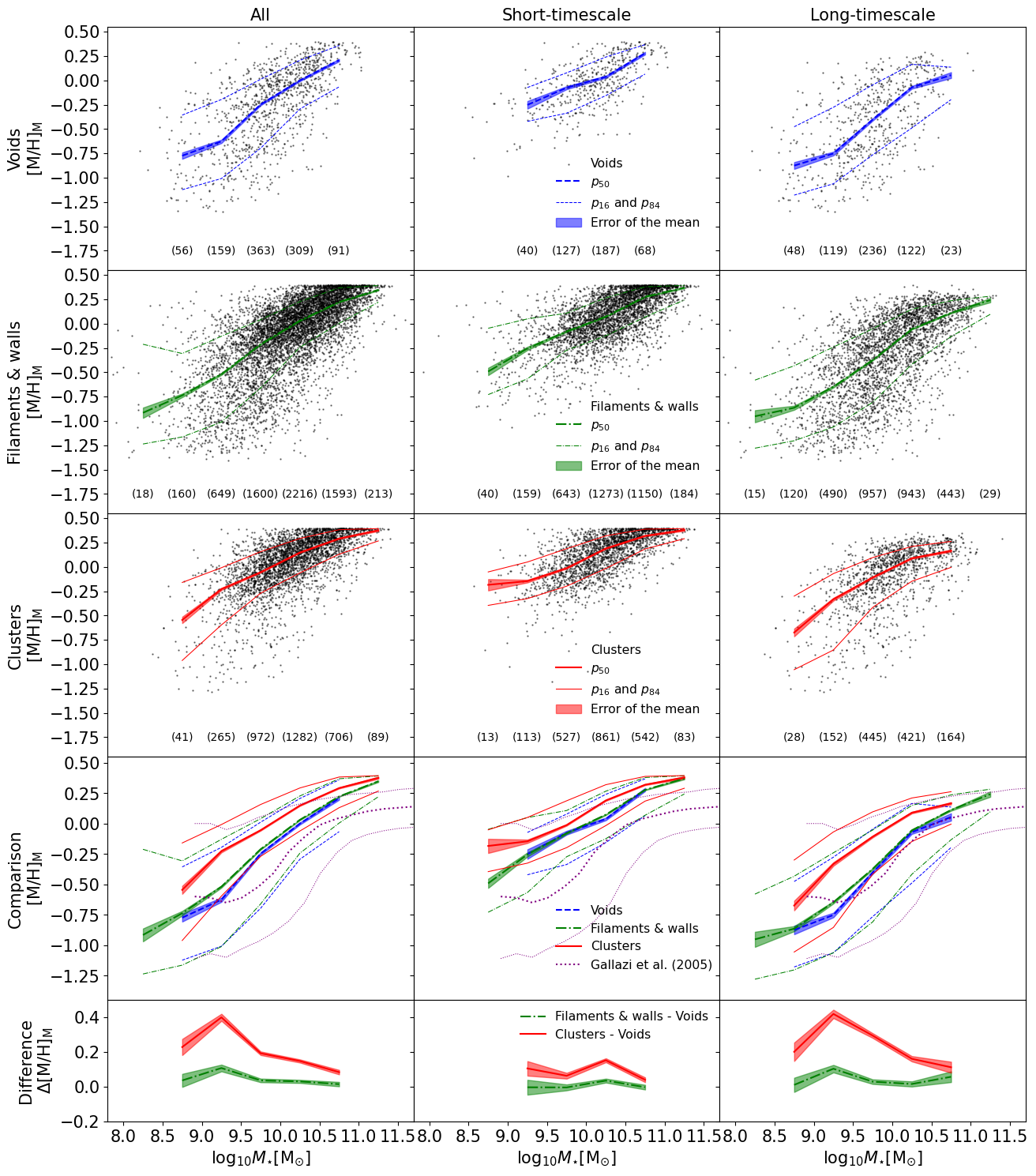}
    \caption{Stellar mass-metallicity relation ($\rm MZ_\star R$). Mass-weighted average stellar metallicity ($\rm[M/H]_M$) as a function of the total stellar mass for the galaxies regardless of their SFH type (left column), galaxies with ST-SFHs (centre column), and galaxies with LT-SFHs (right column) in voids (first row), filaments \& walls (second row), and clusters (third row). The $\rm MZ_\star R$ (dashed blue lines for voids, dot-dashed green lines for filaments \& walls, and solid red lines for clusters) is derived as the 50th percentile (thick lines) inside each stellar mass bin of 0.5 dex. The number of galaxies inside each stellar mass bin is shown in brackets at the bottom of the panels. The shaded areas represent the s.e.m., and the 16th and 84th percentiles (thin lines) show the dispersion of the $\rm MZ_\star R$. The fourth row shows a comparison of the $\rm MZ_\star R$ between galaxies voids, filaments \& walls, and clusters, together with the $\rm MZ_\star R$ from \cite{2005MNRAS.362...41G} as a reference. In the fifth row, we show the differences of the $\rm MZ_\star R$ (lines), together with the error of the difference (shaded areas). See the values reported in tables~\ref{tab:masmetrel_QC} and \ref{tab:masmetdiff_QC}.}
    \label{fig:massmetrel_QC}
\end{figure*}

We show in Figure~\ref{fig:massmetrel_QC} (left column) the $\rm MZ_\star R$ for all the galaxies regardless of their SFH type in voids (first row), filaments \& walls (second row), and clusters (third row). We defined total stellar mass bins of 0.5~dex from $10^{8.0}$ to $10^{11.5}~{\rm M_\odot}$, and obtaine the $\rm MZ_\star R$ (thick lines) as the 50th percentile of the distribution of galaxies inside each stellar mass bin. We estimated the error of the $\rm MZ_\star R$ (shaded areas) as the standard error of the mean (s.e.m.) inside each stellar mass bin. Additionally, we calculated the 16th and the 84th percentiles (thin lines) to visualise the dispersion of the values. We compare the $\rm MZ_\star R$ between the three environments in the fourth row, together with the $\rm MZ_\star R$ from \cite{2005MNRAS.362...41G} as a reference. In the fifth row, we show the $\rm MZ_\star R$ differences between galaxies in voids and filaments \& walls, and also between galaxies in voids and clusters. In Table~\ref{tab:masmetrel_QC} we report the 50th (together with the s.e.m.), 16th, and the 84th percentiles of the $\rm MZ_\star R$ for the different large-scale environments and SFH types. We report in Table~\ref{tab:masmetdiff_QC} the differences of the 50th percentile between voids and filaments \& walls, and also between voids and clusters. 

 In the left column panels of Figure~\ref{fig:massmetrel_QC}, galaxies in voids have slightly lower stellar metallicities on average than galaxies in filaments \& walls, and they are much lower than galaxies in clusters for any given stellar mass, regardless of their SFH type. These differences are more significant at low stellar masses than at high stellar masses, at which the difference might have been diluted by the effect of the metallicity saturation (see Section~\ref{sec:ANA_met}). The stellar metallicity in void galaxies is slightly lower than in filaments \& and walls by \makebox{$0.108\pm0.019$} \makebox{($5.7~\sigma$)} at low stellar masses \makebox{($\sim10^{9.25}~{\rm M_\odot}$)} to \makebox{$0.031\pm0.009$} \makebox{($3.4~\sigma$)} at intermediate stellar masses \makebox{($\sim10^{10.25}~{\rm M_\odot}$).} Void galaxies have lower stellar metallicities than cluster galaxies by \makebox{$0.40\pm0.02$} \makebox{($20.0~\sigma$)} at low stellar masses \makebox{($\sim10^{9.25}~{\rm M_\odot}$)} to \makebox{$0.084\pm0.013$} \makebox{($6.5~\sigma$)} at high stellar masses \makebox{($\sim10^{10.75}~{\rm M_\odot}$).} Our results at very low \makebox{($\sim10^{8.25}~{\rm M_\odot}$)} or very high \makebox{($\sim10^{11.25}~{\rm M_\odot}$)} stellar masses are not statistically significant due to the low number of galaxies in voids (4) and clusters (2). In our conclusions, we only consider stellar mass bins with more than ten galaxies.

 The $\rm MZ_\star R$ derived by \cite{2005MNRAS.362...41G} is in general below what we obtain. They derived the stellar metallicity from the SDSS-DR2 spectra in the centre (3~arcesc aperture) of 175128 galaxies. They used the same type of spectral data as we do in this paper. However, the redshift range of their galaxy sample \makebox{($0.005<z<0.22$)} is much wider than ours \makebox{($0.01<z<0.05$)}. The apparent size of their galaxies is smaller than ours on average. Therefore, the spectrum of their galaxies is integrated over more external regions, where the stellar populations are younger and metal poorer. The spectra of our galaxies are integrated in smaller regions from the centre, where the stars are older and metal richer. At low stellar masses \makebox{($<10^{9.5}~{\rm M_\odot}$)}, our $\rm MZ_\star R$ is similar to theirs. This might be due to the completeness limit of the sample because the number of low-mass galaxies decreases with redshift. Therefore, the low-mass galaxies in their sample might be at similar redshifts as low-mass galaxies in our samples, and the aperture effect explained above is negligible.

\subsection{Effect of the star formation history}
\label{sec:sfh_eff}
In Figure~\ref{fig:massmetrel_QC} (centre and right columns panels) we show how the SFH type (ST-SFH and LT-SFH, respectively) of the galaxies affects the $\rm MZ_\star R$ and compare galaxies in voids, filaments \& walls, and clusters. We find that ST-SFH galaxies have higher stellar metallicities than LT-SFH galaxies for a given stellar mass in the three environments. Together with the fact that it is more likely for galaxies in voids to have an LT-SFH ($51.7\pm0.9\%$) than for galaxies in filaments \& walls ($44.5\pm0.3\%$) and cluster ($36.1\pm0.5\%$), this can explain that void galaxies have slightly lower metallicities on average than galaxies in filaments \& walls, and they are much lower than galaxies in clusters when we compare all the galaxies regardless of their SFH type in Figure~\ref{fig:massmetrel_QC} (left column panels).

We now analyse the stellar metallicity differences for two different SFH types. Galaxies with ST-SFHs have similar stellar metallicities (within the errors) in voids and filaments \& walls, except for intermediate stellar masses \makebox{($\sim10^{10.25}~{\rm M_\odot}$)}, where void galaxies have slightly lower stellar metallicities by \makebox{$0.035\pm0.011$} \makebox{($3.2~\sigma$)}. Galaxies with ST-SFHs in voids have lower stellar metallicities than galaxies in clusters by \makebox{$0.152\pm0.011$} \makebox{($13.8~\sigma$)} at \makebox{$M_\star\sim10^{10.25}~{\rm M_\odot}$} or by \makebox{$0.063\pm0.017$} \makebox{($3.7~\sigma$)} at \makebox{$M_\star\sim10^{9.75}~{\rm M_\odot}$.} Galaxies with LT-SFHs have similar stellar metallicities (within the errors) in voids and filaments \& walls, except for low stellar masses \makebox{($\sim10^{9.25}~{\rm M_\odot}$)}, where void galaxies have slightly lower stellar metallicities by \makebox{$0.104\pm0.021$} \makebox{($5.0~\sigma$)}. Galaxies with LT-SFHs in voids have lower stellar metallicities than in clusters. These differences are more significant at low stellar masses \makebox{($\sim10^{9.25}~{\rm M_\odot}$)}, by \makebox{$0.419\pm0.024$} \makebox{($17.5~\sigma$)} than at high stellar masses \makebox{($\sim10^{10.75}~{\rm M_\odot}$),} by \makebox{$0.11\pm0.03$} \makebox{($3.7~\sigma$).} The differences that we find for galaxies with LT-SFHs are similar to what we find for all the galaxies regardless of their SFH type in Figure~\ref{fig:massmetrel_QC} (left column panels). However, the differences that we find for galaxies with ST-SFHs are less significant. This means that the stellar metallicity differences that we find between galaxies in different environments, regardless of their SFH type, are mainly due to the galaxies with LT-SFHs, while the contribution of galaxies with ST-SFH is not significant.

\subsection{Morphology effect}

\begin{figure*}

    \centering
    \includegraphics[width=0.8\linewidth]{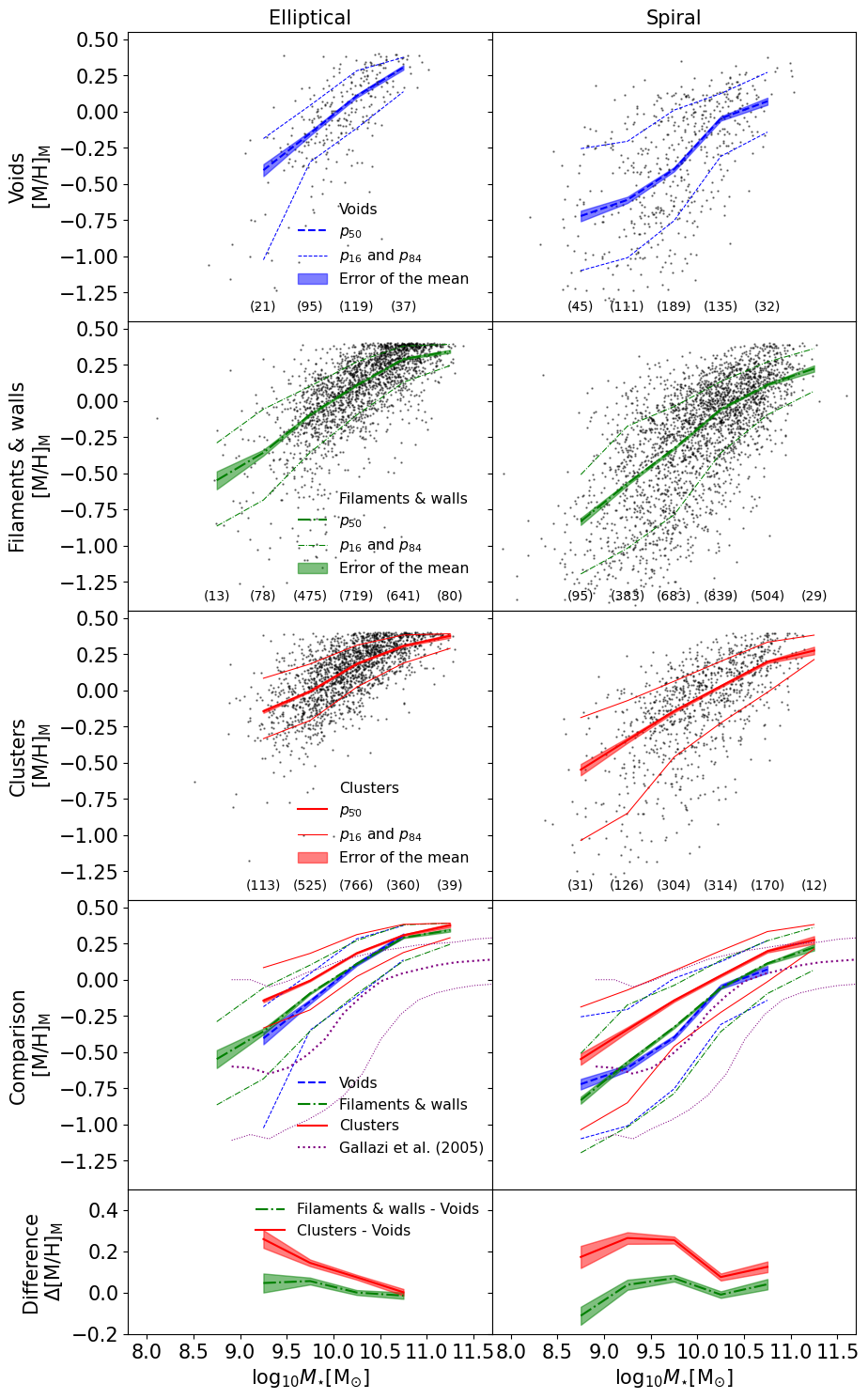}
    \caption{Same as Figure \ref{fig:massmetrel_QC}, but for elliptical (left column) and spiral galaxies (right column). See the values reported in tables \ref{tab:masmetrel_QC_mor} and \ref{tab:masmetdiff_QC_mor}.}
    \label{fig:massmetrel_QC_mor}
\end{figure*}

In Figure~\ref{fig:massmetrel_QC_mor} (left and right columns panels) we analyse the $\rm MZ_\star R$ for different morphological types (elliptical and spiral, respectively) and compare galaxies in voids, filaments \& walls, and clusters. Additionally, we report in tables~\ref{tab:masmetrel_QC_mor} and \ref{tab:masmetdiff_QC_mor} the percentiles (50th with the standard error of the mean (s.e.m.), 16th, and 84th) of the $\rm MZ_\star R$, and the 50th percentile differences between galaxies located in the three large-scale environments. We find that elliptical galaxies have higher stellar metallicities than spiral galaxies in all environments and at all stellar masses. However, galaxies have similar stellar metallicities (within the errors) in voids and filaments \& walls for the two morphological types and for any given stellar mass, except for intermediate stellar masses \makebox{($\sim10^{9.75}~{\rm M_\odot}$)}, where void galaxies have slightly lower stellar metallicities by \makebox{$0.057\pm0.017$} \makebox{($3.4~\sigma$)} for elliptical galaxies and \makebox{$0.071\pm0.016$} \makebox{($4.4~\sigma$)} for spiral galaxies. Void galaxies have lower stellar metallicities than cluster galaxies in the two morphological types. These differences are more significant at low stellar masses \makebox{($\sim10^{9.25}~{\rm M_\odot}$),} by \makebox{$0.26\pm0.04$} \makebox{($6.5~\sigma$)} for elliptical galaxies and \makebox{$0.27\pm0.03$} \makebox{($9.0 ~\sigma$)} for spiral galaxies than at high stellar masses \makebox{($\sim10^{10.25}~{\rm M_\odot}$),} by \makebox{$0.075\pm0.012$} \makebox{($6.2~\sigma$)} for elliptical galaxies and \makebox{$0.077\pm0.017$} \makebox{($4.5~\sigma$)} for spiral galaxies. 

The differences that we find for elliptical galaxies, and also for spiral galaxies are similar to what we find for galaxies with LT-SFHs in Figure~\ref{fig:massmetrel_QC} (right column panels). This occurs more clearly for spiral than for elliptical galaxies. The differences in stellar metallicity between the three environments for spiral galaxies are similar to what we find for LT-SFHs due to the fraction of spiral galaxies with LT-SFHs \citep[62.7\% in voids, 56.5\% in filaments \& walls, and 52.6\% in clusters; see][and Extended Data Figure~7 therein]{2023Natur.619..269D}. A similar effect is visible for elliptical galaxies, which also have similar differences in stellar metallicity between environments as LT-SFH galaxies, but are much more diluted by the higher fraction of elliptical galaxies with ST-SFHs (56.8\% in voids, 65.8\% in filaments \& walls, and 70.9\% in clusters).

\subsection{Colour effect}

\begin{figure*}

    \centering
    \includegraphics[width=0.8\linewidth]{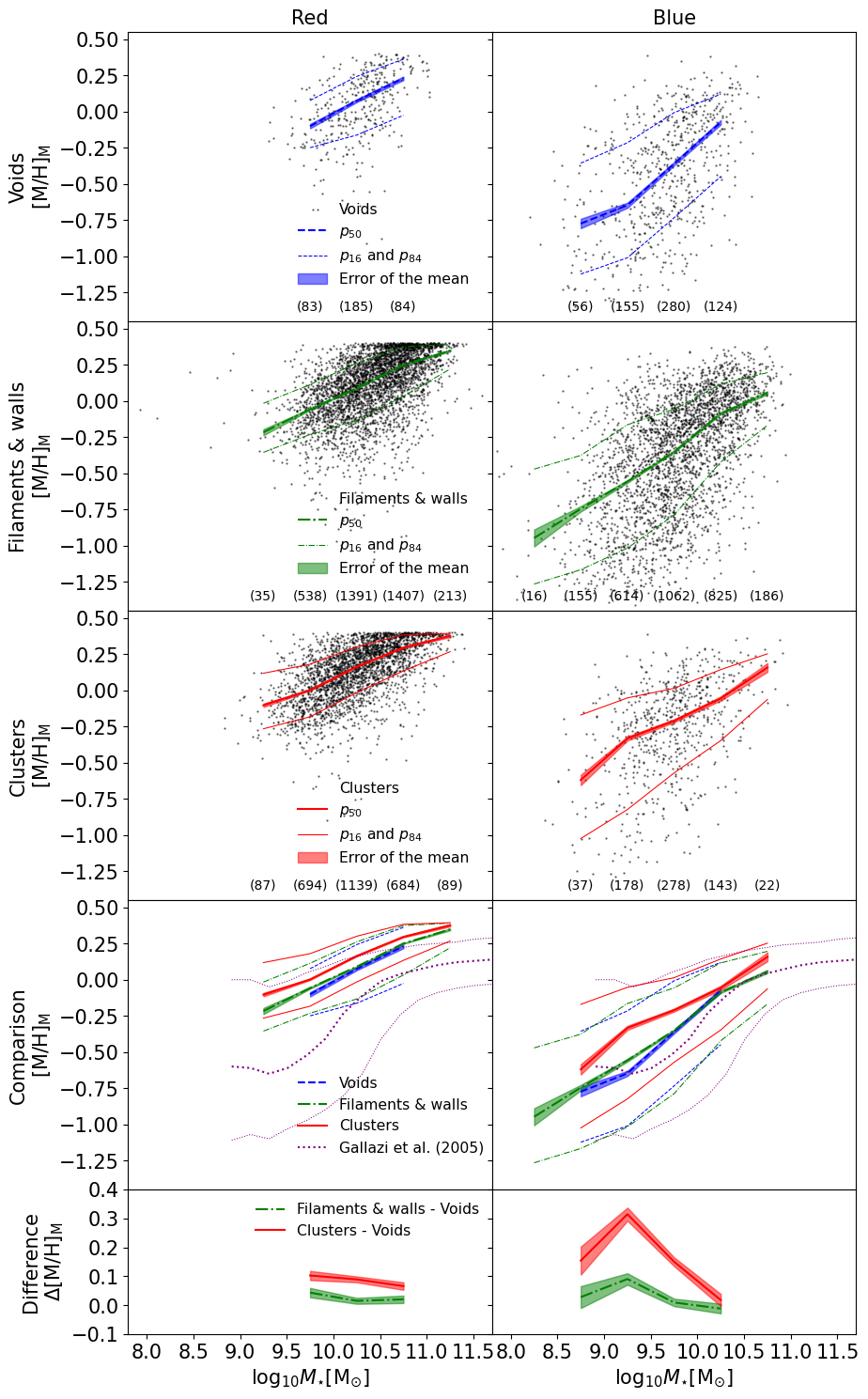}
    \caption{Same as Figure \ref{fig:massmetrel_QC}, but for red (left column) and blue galaxies (right column). See the values reported in tables \ref{tab:masmetrel_QC_col} and \ref{tab:masmetdiff_QC_col}.}
    \label{fig:massmetrel_QC_col}
\end{figure*}

In Figure~\ref{fig:massmetrel_QC_col} we analyse the relation between colour and the $\rm MZ_\star R$, and compare galaxies in voids, filaments \& walls, and clusters. The differences between environments are reported in tables~\ref{tab:masmetrel_QC_col} and \ref{tab:masmetdiff_QC_col}. We find that red galaxies (left column panels) have higher stellar metallicities than blue galaxies (right column panels) for a given stellar mass. However, galaxies have similar stellar metallicities (within the errors) in voids and filaments \& walls for the two colours and for any given stellar mass, except for low stellar masses \makebox{($\sim10^{9.25}~{\rm M_\odot}$)}, where blue galaxies in voids have slightly lower stellar metallicities than in filaments \& walls by \makebox{$0.091\pm0.020$} \makebox{($4.5~\sigma$).} Red galaxies in voids have slightly lower stellar metallicities than in clusters by \makebox{$0.103\pm0.016$} \makebox{($6.4~\sigma$)} at intermediate stellar masses \makebox{($\sim10^{9.75}~{\rm M_\odot}$)} to \makebox{$0.066\pm0.013$} \makebox{($5.1~\sigma$)} at high stellar masses \makebox{($\sim10^{10.75}~{\rm M_\odot}$)}. Blue galaxies in voids have lower stellar metallicities than in clusters by \makebox{$0.315\pm0.023$} \makebox{($13.7~\sigma$)} at low stellar masses \makebox{($\sim10^{9.25}~{\rm M_\odot}$)} to \makebox{$0.150\pm0.015$} \makebox{($10.0~\sigma$)} at intermediate stellar masses \makebox{($\sim10^{9.75}~{\rm M_\odot}$).} 

The stellar metallicity differences of blue galaxies (see the right column panels in Figure~\ref{fig:massmetrel_QC_col}) look very similar to what we find for galaxies with LT-SFH (right column panels in Figure~\ref{fig:massmetrel_QC}), but in a narrower stellar mass range. Blue galaxies in voids have lower stellar metallicities than in clusters for stellar masses between \makebox{$10^{8.75}~{\rm M_\odot}$} and \makebox{$10^{9.75}~{\rm M_\odot}$,} for LT-SFHs galaxies, these differences between voids and clusters remain up to \makebox{$10^{10.25}~{\rm M_\odot}$.} In contrast, the differences in stellar metallicity for red galaxies are less significant, similar to what we find for ST-SFH. The reason is that a high fraction of red galaxies have an ST-SFH and a high fraction of blue galaxies have an LT-SFH \citep[\makebox{$\sim~65-70\%$,} see ][and Extended Data Figure~7 therein]{2023Natur.619..269D}. 

The comparison of blue galaxies between different large-scale environments is not straightforward because the colour distribution is not the same. This also holds for the comparison of red galaxies. This means that a different classification criterion between blue and red galaxies may change our results. However, in Figure~\ref{fig:massmetrel_QC_col0.6} we show the same as in Figure~\ref{fig:massmetrel_QC_col}, but with the colour classification criterion at 0.6 instead of 0.7, for which the metallicity differences that we find are is very similar.

\section{Discussion}\label{sec:discussion}

\subsection{Stellar mass-metallicity relation}

The $\rm MZ_\star R$ correlates with the large-scale environment, SFH type, morphology, and colour of the galaxies. The stellar metallicity of galaxies in voids is slightly lower than in filaments \& walls in specific stellar mass bins, and it is much lower that in clusters at any given stellar mass. Many works have studied the $\rm MZ_g R$ of galaxies, and some even compared different local \citep{2011AJ....141..162D, 2012MNRAS.425..273P, 2017MNRAS.465.1358P} and large-scale environments \citep{2008AJ....136....1W, 2011AstBu..66..255P, 2015ApJ...798L..15K,  2019ApJ...883...29W}, but there is no consensus about the gas metallicity properties in void galaxies. However, only a few works studied how the local environment affects the $\rm MZ_\star R$, and none of them studied how it is affected by the large-scale structures of the Universe. 

\cite{2010MNRAS.407..937P} and \cite{2021MNRAS.502.4457G} studied the stellar populations of galaxies in groups and found that the stellar age and metallicity of central galaxies increases with the halo mass when they averaged for all stellar masses. However, \cite{2022MNRAS.511.4900S} found that this correlation is produced by the stellar-to-halo mass relation. The stellar age and metallicity of central galaxies decreases with halo mass for a fixed stellar mass in low-mass haloes ($<10^{13.5}~{\rm M_\odot}$). This means that the lower stellar metallicities that we find for void galaxies might be due to their higher halo-to-stellar mass ratios compared to denser environments \citep{2018MNRAS.480.3978A, 2020A&A...638A..60A, 2020MNRAS.493..899H, 2022MNRAS.tmp.2385R}. However, they did not compare with void galaxies. 

\cite{2020A&A...639A..71K} studied the effect of the local and large-scale environment on the properties of galaxies in groups (i.e. from two to six galaxies), such as colour, stellar mass, morphology, and the 4000~\AA~break. They compared galaxies inside (within a radius of 1~Mpc from the filament axes) and outside the filament (within a radius of 1~Mpc to 4~Mpc) and concluded that the effect of filaments on the properties of galaxies in groups is marginal, and that the local environment is the main factor that determines their properties. However, the delimitation of voids within the large-scale structure of the Universe and the classification of void galaxies is not as easy as measuring their distance to the closest filament, and many of the galaxies outside the filament might be inside a wall, whose galaxies have similar properties as galaxies inside the filaments. Furthermore, they did not study the galaxies in the deepest regions of the voids, which may be located up to 15~Mpc from the closest filament. Clearly, the local density is lower in voids than in filaments on average, but both the local and large-scale environments influence the properties of galaxies directly or indirectly.

\cite{2021MNRAS.502.4457G} found that the stellar metallicity of recent cluster or group infallers (i.e. galaxies that passed the virial radius of the host halo <~2.5~Gyr ago) is lower than that of those that have been exposed to the environment of clusters or groups for a longer time (>~2.5~Gyr). Recent infallers continue to form stars (although at lower rates) for \makebox{$\sim$~2~Gyr} after their infall \citep[][Figure~7 therein]{2020ApJS..247...45R}. The typical timescale of ram-pressure stripping in clusters as massive as Virgo is <~1~Gyr, and \makebox{$\sim$~2~Gyr} for gas stripping due to tidal interactions \citep{2006PASP..118..517B}. This means that for about 2~Gyr after falling into the cluster, the galaxies (recent infaller) still act as if they were in the previous host environment (i.e. voids or filaments \& walls) with little changes in the SFR. \cite{2021MNRAS.502.4457G} results would agree with ours because they found that recent infallers (i.e. with void and filament \& wall galaxy properties) have lower stellar metallicities than ancient infallers (with cluster galaxy properties).

It is well known that the metallicity of a galaxy is significantly determined by its SFH \citep{2002A&A...388..396T}, and this is reflected in our results. The stellar metallicity differences that we find for galaxies in different large-scale environments are much more significant for LT-SFH than for ST-SFH. ST-SFH galaxies formed a high fraction \makebox{($\sim27\%$)} of their stellar mass very early \citep[more than 12.5~Gyr ago][]{2023Natur.619..269D}, enriching their ISM very quickly for the next generation of stars to be formed. Furthermore, they assembled 50\% of their stellar mass at a similar time \makebox{($\sim$~11~Gyr ago)} in voids, filaments \& walls, and clusters. This suggests that in the early Universe, the contrast between the large-scale environments was lower, which did not affect the evolution of the ST-SFH galaxies in the beginning but later, that is, when they formed 70\% of their stellar mass. Void galaxies assembled 70\% of their stellar mass later than in filaments \& walls \makebox{(by $\sim1~{\rm Gyr}$)} and much later than in clusters \makebox{(by $\sim2~{\rm Gyr}$)}, more significantly at low \makebox{($10^{9.0}-10^{9.5}~{\rm M_\odot}$)} than at high stellar masses \makebox{($10^{10.0}-10^{10.5}~{\rm M_\odot}$)}. 

\begin{figure}[!h]
    \centering
    \begin{subfigure}[b]{\linewidth}
        \centering
        \includegraphics[width=\linewidth]{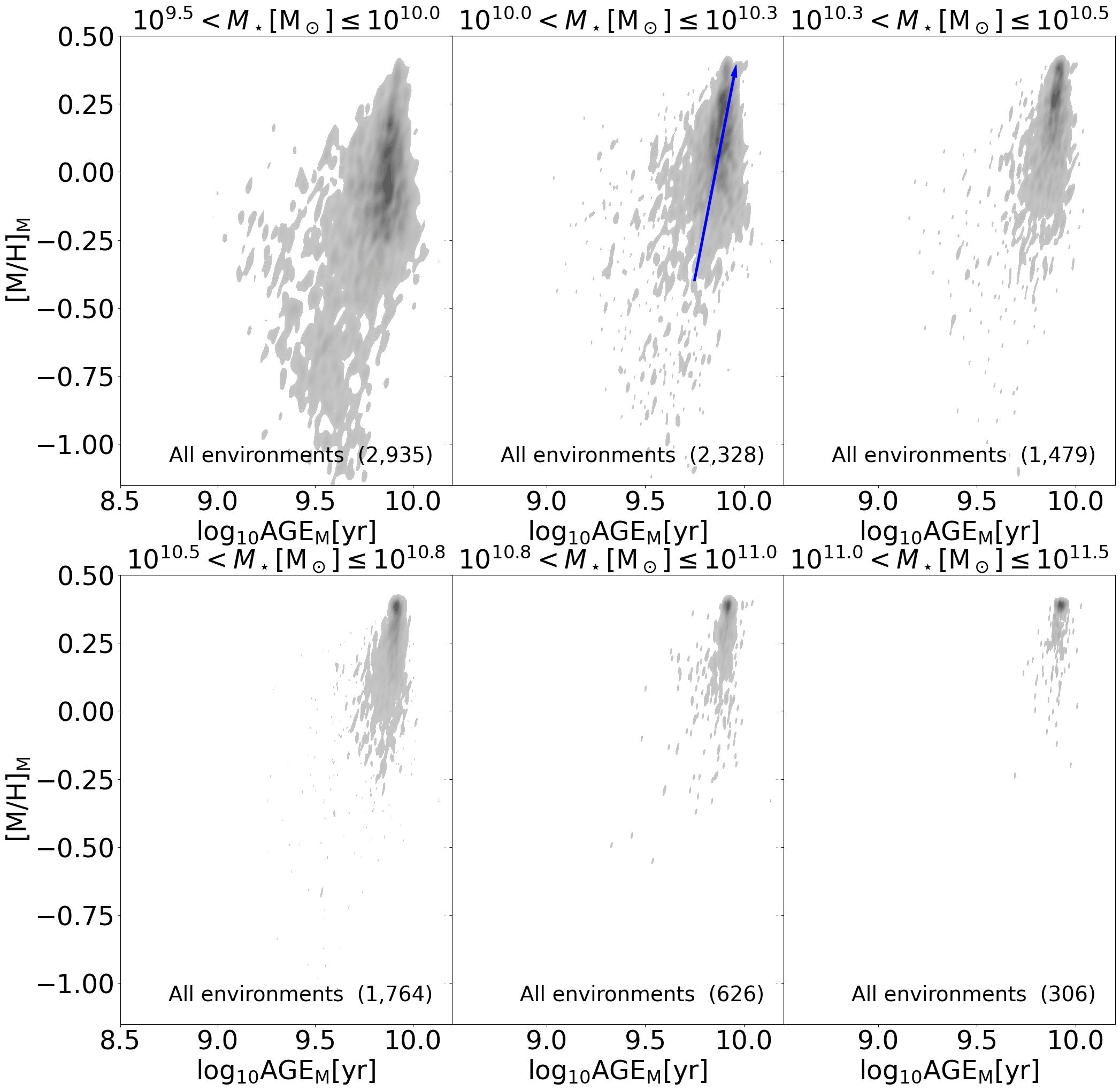}
        \caption{Mass-weighted}
        \label{fig:age_met_M}
    \end{subfigure}
    \hfill
    \begin{subfigure}[b]{\linewidth}
        \centering
        \includegraphics[width=\linewidth]{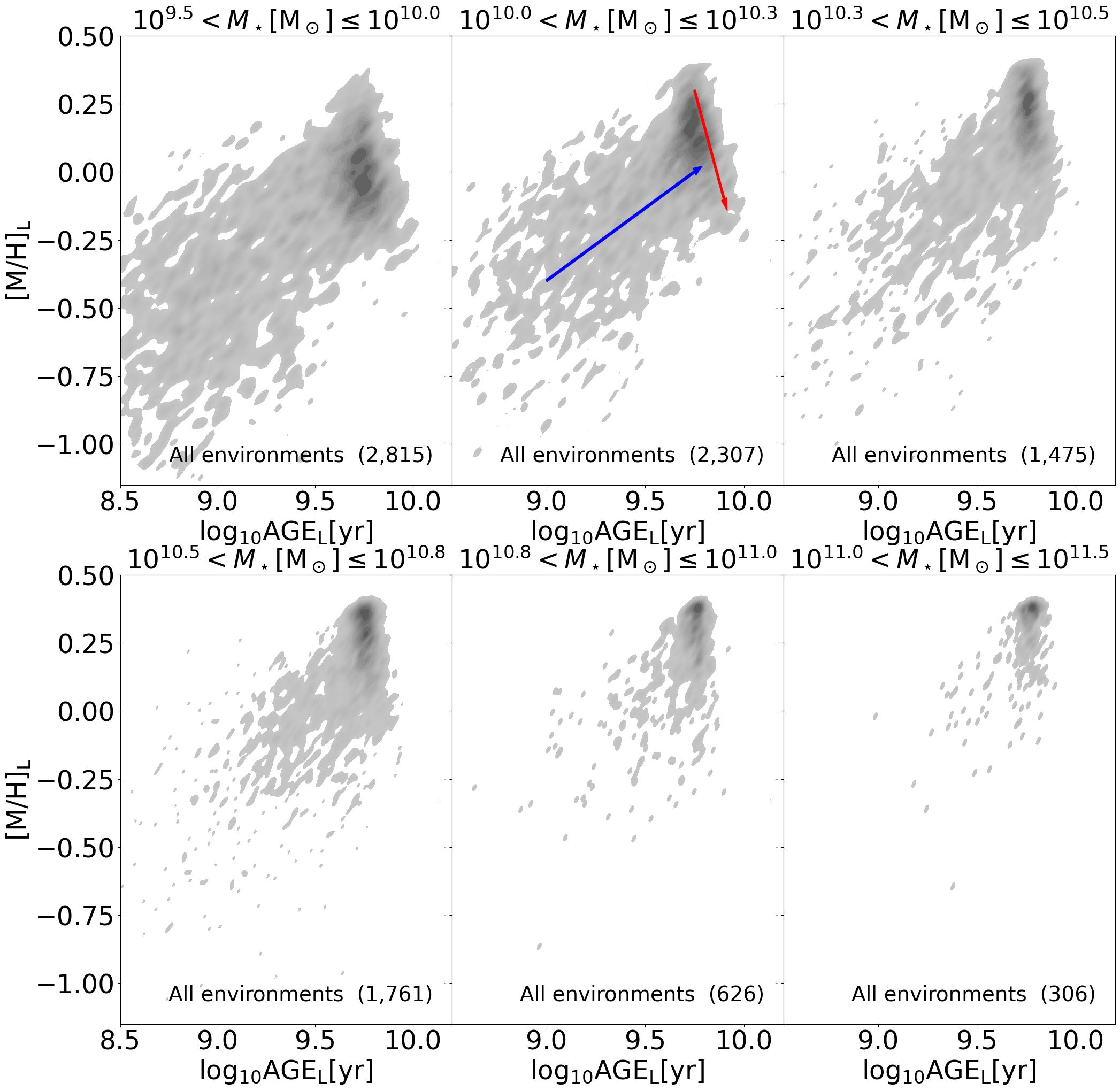}
        \caption{Luminosity-weighted}
        \label{fig:age_met_L}
    \end{subfigure}
    \caption{Stellar age-metallicity distribution for several stellar mass ranges. The stellar metallicities and ages of the galaxies are derived as the mass-weighted (a) and luminosity-weighted (b) averages. The arrows illustrate the age-metallicity correlation (blue arrows) and anti-correlation (red arrow).}
    \label{fig:age_met}
\end{figure}

The LT-SFH galaxies have had a more steady SFHs, enriching their ISM more slowly, and possibly diluting their metallicity by metal-poor gas accretion. Moreover, LT-SFH galaxies have been affected by their large-scale environments since very early on, assembling their stellar mass later in voids than in filaments \& walls (at intermediate stellar masses, \makebox{$10^{9.5}-10^{10.0}~{\rm M_\odot}$,} by \makebox{$\sim1~{\rm Gyr}$)}, and much later than in clusters (at any given stellar mass by \makebox{$\sim2~{\rm Gyr}$).} This indicates that galaxies that had similar SFHs in the beginning (i.e. ST-SFH galaxies that assembled 50\% of their stellar mass at similar times in the three environments) would have similar stellar metallicities now, even if their SFH differ later (assembling 70\% of their stellar mass at different times). In contrast, galaxies with different SFHs in the beginning (i.e. LT-SFH galaxies that assembled 50\% of their stellar mass at different times in the three environments) would have different stellar metallicities now. Therefore, the mass-weighted stellar metallicity of a galaxy is mostly determined by its initial SFH period (old stars). We confirm this in Figure~\ref{fig:age_met_M}, where we show a direct correlation (blue arrow) in the mass-weighted stellar age-metallicity distribution. We derived the mass-weighted stellar age following the same recipe as for metallicity given in \makebox{Equation~\ref{eq:MW-met}.}

\cite{2005MNRAS.362...41G} found an effect of the age-metallicity degeneracy as an anti-correlation for the oldest galaxies in the stellar age-metallicity distribution for luminosity-weighted ages and metallicities (see their figures~11 and 12). We do not find this anti-correlation for the mass-weighted average in Figure~\ref{fig:age_met_M}. However, the comparison between luminosity-weighted and mass-weighted averages is not straightforward. We derive
in Appendix~\ref{sec:MZLR} the $\rm MZ_\star R$ applying the luminosity-weighted average, where we find similar results to what we find in Section~\ref{sec:results} for the mass-weighted average. We show in Figure~\ref{fig:age_met_L} the luminosity-weighted stellar age-metallicity distribution for our sample of galaxies, and we recover a similar anti-correlation (red arrow) for the oldest galaxies. However, the direct correlation (blue arrow) for young galaxies is maintained. Void galaxies have assembled their mass more slowly on average than galaxies in denser environments \citep[see ][]{2023Natur.619..269D}, and their stellar populations are consequently younger than in filaments \& walls, and clusters. This agrees with the lower stellar metallicities that we find for void galaxies.

We also show that the stellar metallicity differences that we find between different large-scale environments are more significant for blue spiral galaxies than for red elliptical galaxies. \cite{2005MNRAS.362...41G} found that for a given stellar mass, early-type galaxies have older and more metal-rich stellar populations on average than late-type galaxies. Red elliptical galaxies are more likely to be gas-poor quenched massive galaxies that have suffered high star formation bursts after galaxy-galaxy interactions or mergers, quickly enriching their ISM. They also suffer from internal feedback processes such as AGNs, supernovae, and stellar winds. Blue spiral galaxies are more likely to be gas-rich star-forming galaxies with smaller chances to have suffered mergers in the past, preserving their metal-poor surrounding gas, and enriching their ISM slower than red elliptical galaxies. Void galaxies are bluer on average and have later morphological types than in denser environments, fulfilling the expectations of finding lower stellar metallicities in void galaxies.

\subsection{Scatter around the gas-phase mass-metallicity relation}

As very little is known about the correlation between the $\rm MZ_\star R$ and the large-scale environment, we compared our results with previous studies about the gas-phase metallicity. It is not obvious to directly compare the $\rm MZ_\star R$ with studies focusing on the $\rm MZ_g R$ as this procedure only takes into account star-forming galaxies (with gas emission lines), but we also considered quenched galaxies (without gas emission lines). We can only refer to blue and spiral (i.e. star-forming) galaxies of our sample (right column panels in figures~\ref{fig:massmetrel_QC_mor} and \ref{fig:massmetrel_QC_col}, respectively) when comparing our $\rm MZ_\star R$ results with the $\rm MZ_gR$. Additionally, previous studies \citep{2005MNRAS.362...41G, 2008MNRAS.391.1117P, 2017ApJ...847...18Z} have shown that the stellar metallicity correlates with the gas-phase metallicity when considering young stellar populations or the luminosity-weighted average stellar metallicity. We show in Appendix~\ref{sec:MZLR} the $\rm MZ_\star R$ applying the luminosity-weighted average, where we find similar results to what we find in Section~\ref{sec:results} for the mass-weighted average of blue and spiral galaxies. Blue and spiral galaxies in voids have lower stellar metallicities than in filaments \& walls and much lower than in clusters, and the differences are more significant at low than at high stellar mass. Thus, we assume that it is fair to compare our mass-weighted $\rm MZ_\star R$ for blue spiral galaxies with the $\rm MZ_g R$ results in the literature, and we interpret our differences in the $\rm MZ_\star R$ for different large-scale environments as scatter around the main $\rm MZ_g R$, associated with the halo mass of the galaxies, gas accretion, and gas feedback.

IllustrisTNG simulation \citep{2019MNRAS.484.5587T} showed that the scatter in the $\rm MZ_g R$ correlates with the ISM-to-stellar mass ratio ($M_{\rm ISM}/M_\star$) and the SFR. High SFRs or low $M_{\rm ISM}/M_\star$ ratios increase the gas metallicity of star-forming galaxies. This would imply that the higher stellar metallicity that we find for the blue and spiral galaxies in clusters is driven by a lower $M_{\rm ISM}/M_\star$ or higher SFR than in less dense environments.

Observations \citep{2022arXiv221210657Y} show that the gas mass and the gas metallicity are anti-correlated at a given stellar mass up to \makebox{$M_\star<10^{10.5}~{\rm M_\odot}$.} Simulations \citep{2011MNRAS.414.2458V} predict a correlation between the gas-accretion rates and the host halo mass of the galaxy up to \makebox{$M_{\rm h}<10^{12.0}~{\rm M_\odot}$.} This means that the more massive the halo, the higher the gas accretion and the gas fraction. This implies a decreased gas-phase metallicity because the inflowing gas is assumed to be more metal poor than the ISM. This would imply that the lower stellar metallicity that we find for our void galaxies is due to a higher gas mass or higher gas-accretion rates driven by their more massive host haloes. This scenario is supported by simulations that find that the halo-to-stellar mass ratio is higher in void galaxies than in galaxies in denser large-scale environments \citep{2018MNRAS.480.3978A, 2020A&A...638A..60A, 2020MNRAS.493..899H, 2022MNRAS.tmp.2385R}. Another possibility is that there are two different modes of gas accretion \citep{2005MNRAS.363....2K}: the cold gas-accretion mode dominates in void galaxies, while the hot gas-accretion mode prevails in denser environments. Additionally, a previous observational study \citep{2021ApJ...906...97F} found that void galaxies have a higher gas mass than galaxies in denser environments. This is more significant for low-mass \makebox{($M_\star<10^{10.0}~{\rm M_\odot}$)} and early-type galaxies than for high-mass or late-type galaxies. However, other observational studies \citep{1996AJ....111.2150S, 2012AJ....144...16K, 2022A&A...658A.124D} and simulations \citep{2022MNRAS.tmp.2385R} did not find significant gas-mass differences between galaxies in different large-scale environments for this stellar mass regime. Very little observational evidence has been provided for gas-accretion rates onto galaxies, given the considerable challenge of directly measuring it.

At higher stellar masses \makebox{($M_\star>10^{10.5}~{\rm M_\odot}$),} \cite{2022arXiv221210657Y} reported that the scatter of the $\rm MZ_g R$ correlates only very weakly with the gas mass, but a stronger trend is found with AGN activity. Both the EAGLE and IllustrisTNG simulations find that the scatter of the $\rm MZ_g R$ is no longer driven by systematic variations in gas-inflow rate, but is instead dominated by the impact of AGN feedback \citep{2017MNRAS.472.3354D, 2019MNRAS.484.5587T, 2021MNRAS.504.4817V}. Galaxies with a higher nuclear activity have the lowest metallicities. Previous observational studies \citep{2008ApJ...673..715C, 2022MNRAS.509.1805C} found a larger fraction of AGNs or massive black holes (BH) in voids than in denser large-scale environments. However, there is as yet no consensus on the effect of the large-scale structure on the nuclear activity in galaxies, as \cite{2018A&A...620A.113A} found the opposite result for quenched isolated galaxies, and other studies did not find significant differences in the fraction of AGNs \citep[observation]{2019ApJ...874..140A} or in the BH-to-galaxy mass ratio \citep[simulation]{2020MNRAS.493..899H} between different large-scale environments. This would agree with our results because we do not find significant stellar metallicity differences between galaxies in different large-scale environments at high stellar masses \makebox{($M_\star>10^{10.5}~{\rm M_\odot}$).} However, in this stellar mass range, the stellar metallicity differences that we find between galaxies in voids and denser large-scale environments might have been diluted by the effect of the metallicity saturation.

\section{Conclusions}\label{sec:conclusions}

We applied a non-parametric full spectral fitting analysis to the SDSS spectra in the centre of statistically sound samples of galaxies in voids, filaments \& walls, and clusters. We recovered their stellar populations and studied how the large-scale structures of the Universe affect the stellar mass-metallicity relation of galaxies for different SFH types, morphologies, and colours. The main conclusions are listed below.

\begin{enumerate}
               
        \item Void galaxies have slightly lower stellar metallicities than galaxies in filaments \& walls, more significantly so at low stellar masses \makebox{($10^{9.25}~{\rm M_\odot}$} by \makebox{$0.108\pm0.019$)} than at high stellar masses \makebox{($10^{10.25}~{\rm M_\odot}$} by \makebox{$0.031\pm0.009$)}, and much lower than galaxies in clusters, also more significantly so at low stellar masses \makebox{($10^{9.25}~{\rm M_\odot}$} by \makebox{$0.40\pm0.02$)} than at high stellar masses \makebox{($10^{10.75}~{\rm M_\odot}$} by \makebox{$0.084\pm0.013$).} At high stellar masses \makebox{($10^{10.75}~{\rm M_\odot}$),} the differences between galaxies in voids and denser environments might have been diluted by the effect of the metallicity saturation.

        \item The stellar metallicity differences between galaxies in voids and galaxies in denser environments are more significant for LT-SFH than for ST-SFH galaxies. This is more significant for spiral than for elliptical galaxies, and it is more significant for blue than for red galaxies.

        \item  The ST-SFH galaxies in voids have slightly lower stellar metallicities than galaxies in filaments \& walls at intermediate stellar masses \makebox{($10^{10.25}~{\rm M_\odot}$,} by \makebox{$0.035\pm0.011$),} and much lower than galaxies in clusters, more significantly so at intermediate stellar masses \makebox{($10^{10.25}~{\rm M_\odot}$,} by \makebox{$0.152\pm0.011$)} than at low stellar masses \makebox{($10^{9.25}~{\rm M_\odot}$,} by \makebox{$0.063\pm0.017$).} LT-SFH galaxies in voids and filaments \& walls have similar stellar metallicities, except for low stellar masses \makebox{($10^{9.25}~{\rm M_\odot}$),} where void galaxies have slightly lower stellar metallicities (by \makebox{$0.104\pm0.021$).} LT-SFH galaxies in voids have much lower stellar metallicities than in clusters, more significantly so at low stellar masses \makebox{($10^{9.25}~{\rm M_\odot}$,} by \makebox{$0.419\pm0.024$),} than at high stellar masses \makebox{($10^{10.75}~{\rm M_\odot}$,} by \makebox{$0.11\pm0.03$).} 

        \item Both elliptical and spiral galaxies in voids have slightly lower stellar metallicities than in filaments \& walls at intermediate stellar masses \makebox{($10^{9.75}~{\rm M_\odot}$,} by \makebox{$0.057\pm0.017$} and \makebox{$0.071\pm0.016$,} respectively), and much lower than in clusters, more significantly so at low stellar masses \makebox{($10^{9.25}~{\rm M_\odot}$,} by \makebox{$0.26\pm0.04$ and $0.27\pm0.03$,} respectively) than at high stellar masses \makebox{($10^{10.25}~{\rm M_\odot}$,} by \makebox{$0.075\pm0.012$} and \makebox{$0.077\pm0.017$,} respectively).
 
        \item Blue galaxies in voids have slightly lower stellar metallicities than in filaments \& walls at low stellar masses \makebox{($10^{9.25}~{\rm M_\odot}$,} by \makebox{$0.091\pm0.020$)}, but no significant differences are found between red galaxies in voids and filaments \& walls. Red galaxies in voids have slightly lower stellar metallicities than in clusters, more significantly so at lower stellar masses \makebox{($10^{9.75}~{\rm M_\odot}$} by \makebox{$0.103\pm0.016$)} than at higher stellar masses \makebox{($10^{10.75}~{\rm M_\odot}$} by \makebox{$0.066\pm0.013$).} Blue galaxies in voids have lower stellar metallicities than in clusters, more significantly so at lower stellar masses \makebox{($10^{9.25}~{\rm M_\odot}$} by \makebox{$0.315\pm0.023$)} than at higher stellar masses \makebox{($10^{9.75}~{\rm M_\odot}$} by \makebox{$0.150\pm0.015$).}
        
    \end{enumerate}

    In summary, galaxies in voids and filaments \& walls have similar stellar metallicities, except for intermediate stellar masses, where void galaxies have slightly lower stellar metallicities. Void galaxies have lower stellar metallicities than galaxies in clusters (more significantly so at low than at high stellar masses). These differences can be explained by the slower SFHs of voids galaxies, specially at the initial star formation period in LT-SFH galaxies. Additionally, when we assume a good correlation between the stellar and gas-phase metallicities of star-forming galaxies (i.e. blue and spiral galaxies, for which we find the highest stellar metallicity differences between galaxies in different large-scale environments compared to red and elliptical galaxies, where the differences are less significant), we can explain the lower stellar metallicity of void galaxies at intermediate stellar masses \makebox{($M_\star<10^{10.5}~{\rm M_\odot}$)} based on their higher halo-to-stellar mass ratio, and higher gas accretion. However, further research is needed in terms of gas-mass content, gas-phase metallicity, and nuclear activity of void galaxies, together with the correlation between these physical processes and the SFH. The CAVITY project aims to fulfil these needs by analysing the resolved emission lines, stellar populations, and kinematics from PPAK IFU data, together with ancillary data, such as atomic and molecular gas, of galaxies in voids.

\begin{acknowledgements}
We acknowledge financial support by the research projects AYA2017-84897-P, PID2020-113689GB-I00, and PID2020-114414GB-I00, financed by MCIN/AEI/10.13039/501100011033, the project A-FQM-510-UGR20 financed from FEDER/Junta de Andalucía-Consejer\'ia de Transforamción Económica, Industria, Conocimiento y Universidades/Proyecto and by the grants P20\_00334 and FQM108, financed by the Junta de Andaluc\'ia (Spain).
J.F-B.  acknowledges support through the RAVET project by the grant PID2019-107427GB-C32 from the Spanish Ministry of Science, Innovation and Universities (MCIU), and through the IAC project TRACES which is partially supported through the state budget and the regional budget of the Consejer\'ia de Econom\'ia, Industria, Comercio y Conocimiento of the Canary Islands Autonomous Community.
S.D.P. acknowledges financial support from Juan de la Cierva Formaci\'on fellowship (FJC2021-047523-I) financed by MCIN/AEI/10.13039/501100011033 and by the European Union "NextGenerationEU"/PRTR, Ministerio de Econom\'ia y Competitividad under grant PID2019-107408GB-C44, from Junta de Andaluc\'ia Excellence Project P18-FR-2664, and also from the State Agency for Research of the Spanish MCIU through the `Center of Excellence Severo Ochoa' award for the Instituto de Astrof\'isica de Andaluc\'ia (SEV-2017-0709).
G.B-C. acknowledges financial support from grants PID2020-114461GB-I00 and CEX2021-001131-S, funded by MCIN/AEI/10.13039/501100011033, from Junta de Andalucía (Spain) grant P20-00880 (FEDER, EU) and from grant PRE2018-086111 funded by MCIN/AEI/10.13039/501100011033 and by 'ESF Investing in your future'.
R.G-B. acknowledges financial support from the grants CEX2021-001131-S funded by MCIN/AEI/10.13039/501100011033, PID2019-109067-GB100 and to and to CSIC “Ayudas de Incorporación” grant 202250I003.
M.A-F. acknowledges support the Emergia program (EMERGIA20\_38888) from Consejería de Transformación Económica, Industria, Conocimiento y Universidades and University of Granada.

This research made use of Astropy, a community-developed core Python (http://www.python.org) package for Astronomy
    ; ipython
    ; matplotlib
    ; SciPy, a collection of open source software for scientific computing in Python
    ; APLpy, an open-source plotting package for Python
    ; and NumPy, a structure for efficient numerical computation
    .
\end{acknowledgements}    


\clearpage
\begin{appendix}

\section{Sub-samples with the same stellar mass distribution}\label{sec:sample-KS}

\begin{figure*}
        \centering
        \includegraphics[width=\textwidth]{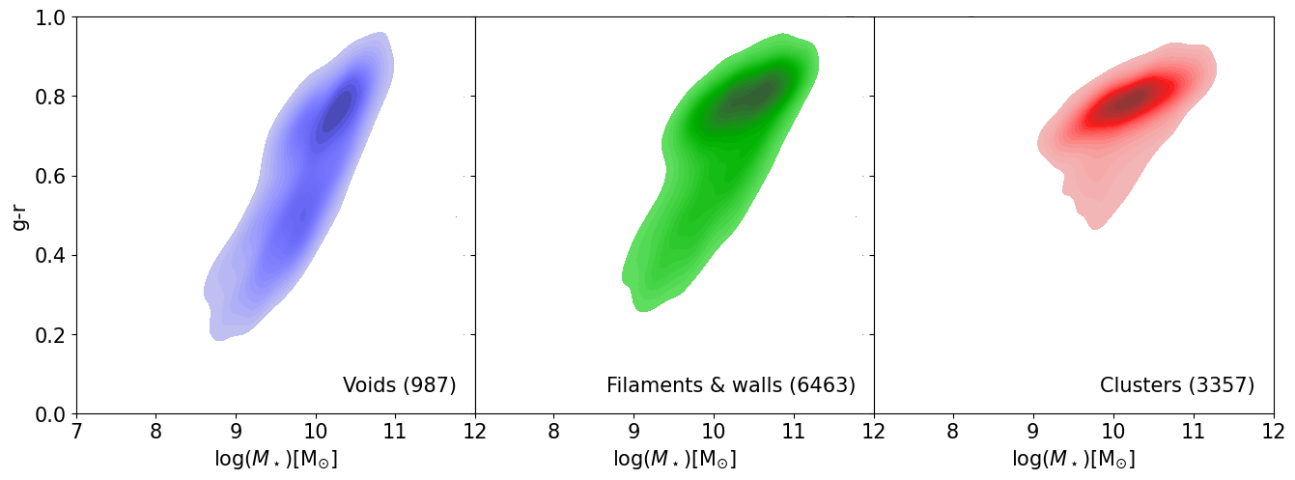}
        \caption{Colour vs. stellar diagram distribution of our samples of galaxies in voids (left), filaments \& walls (centre), and clusters (right). The number of galaxies in each sample is shown between brackets in the legend.}
        \label{fig:col_mass_mask}
    \end{figure*}

We show in Figure~\ref{fig:col_mass_mask} the colour vs. stellar mass diagram distribution of our galaxy samples. Void galaxies are bluer on average and less massive than galaxies in denser environments. Our analysis might be affected by the different stellar mass distribution of the three samples. We defined five stellar mass bins of 0.5~dex between $10^{8.5}$ and \makebox{$10^{11.0}~{\rm M_\odot}$} and selected random sub-samples with the same stellar mass distribution as our void galaxy sample inside each stellar mass bin. The number of galaxies beyond these limits was not enough to define sub-samples with a similar stellar mass distribution applying a KS-test. We were left with 978 galaxies in voids, 4800 galaxies in filaments \& walls, and 2570 galaxies in clusters for our study.

\begin{figure*}

    \centering
    \includegraphics[width=\linewidth]{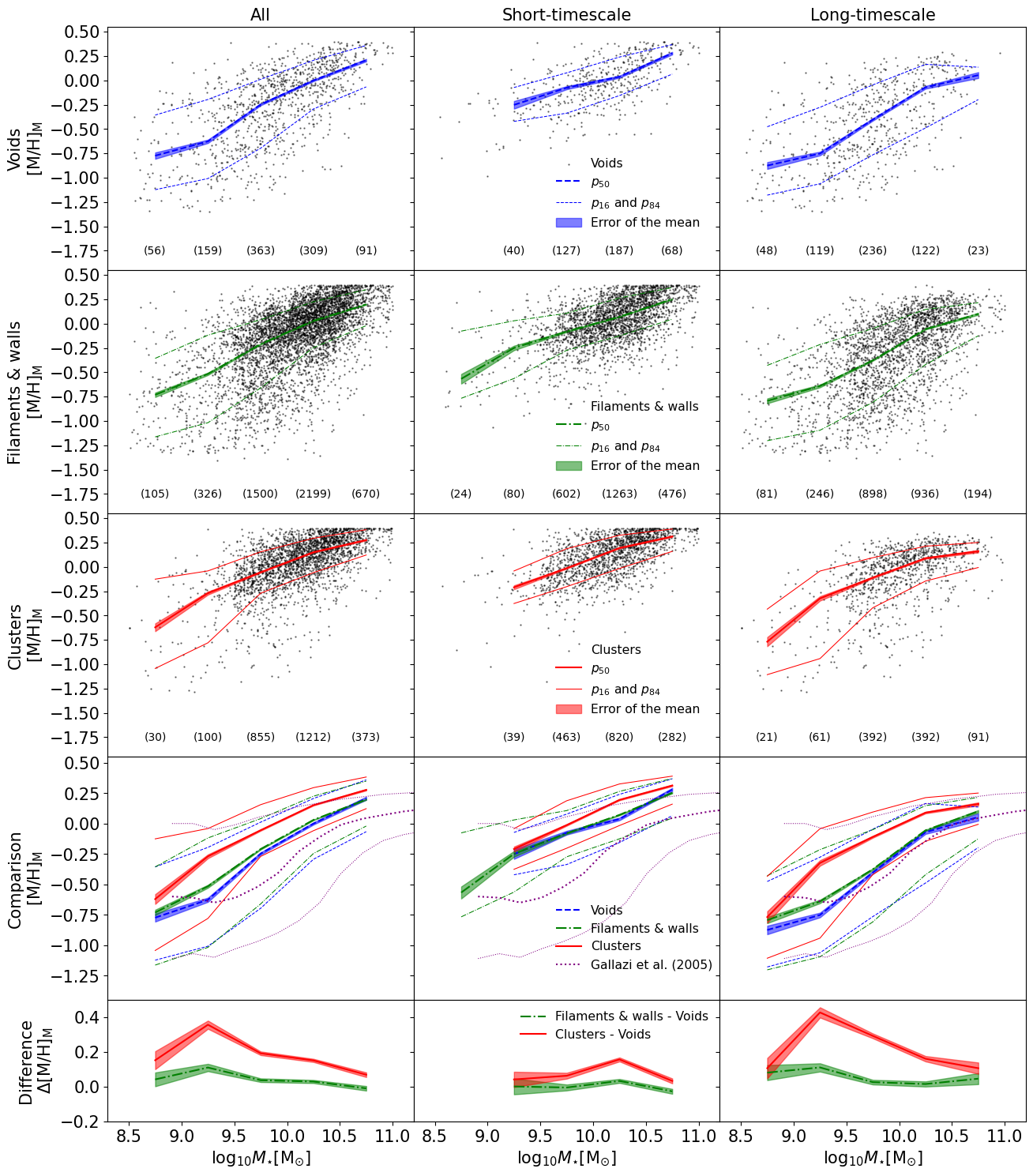}
    \caption{Same as Figure \ref{fig:massmetrel_QC}, but for the sub-samples of galaxies with the same stellar mass distribution. See the values reported in tables \ref{tab:masmetrel_KS} and \ref{tab:masmetdiff_KS}.}
    \label{fig:massmetrel_KS}
\end{figure*}

\begin{figure*}

    \centering
    \includegraphics[width=0.73\linewidth]{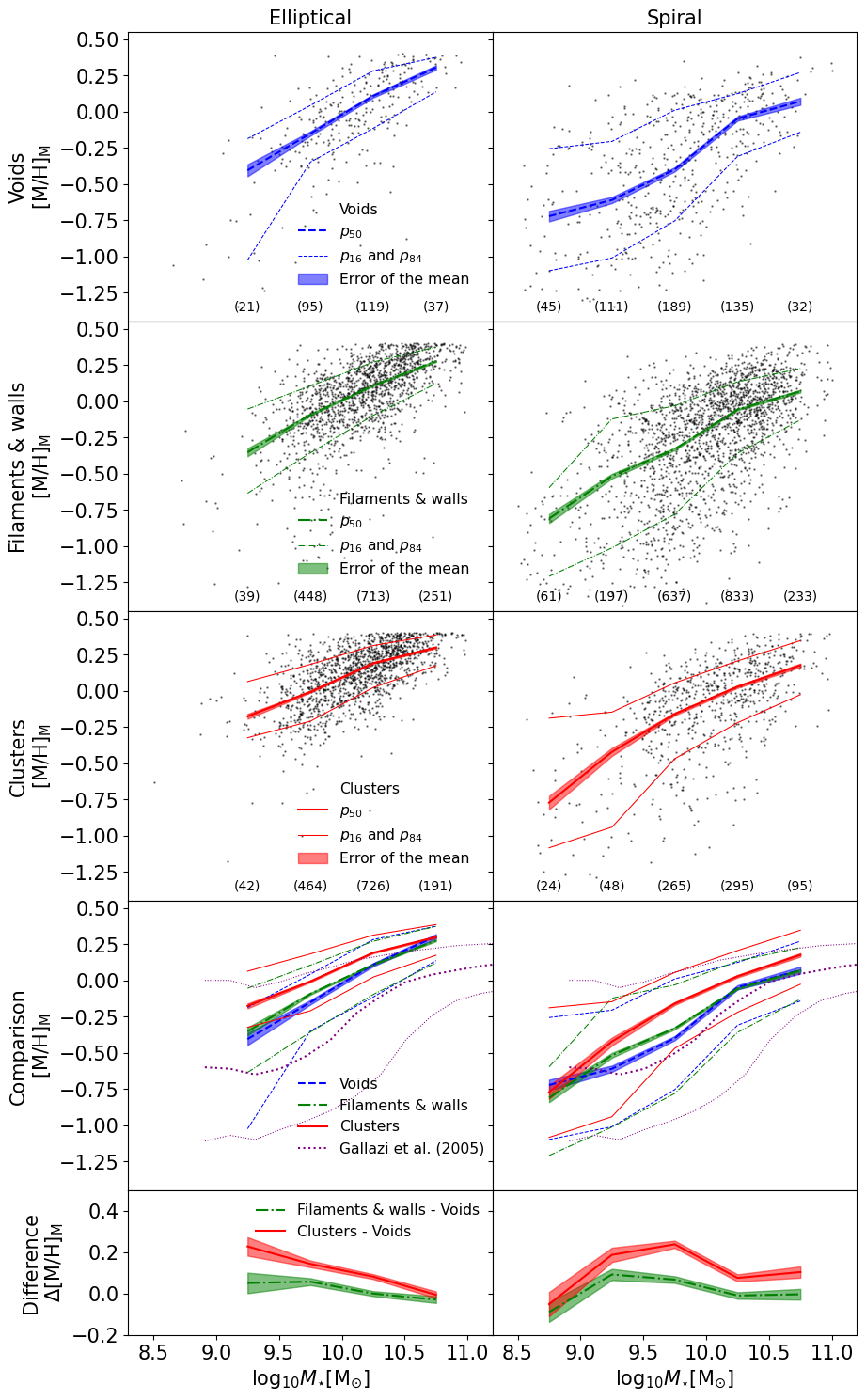}
    \caption{Same as Figure \ref{fig:massmetrel_QC_mor}, but for the sub-samples of galaxies with the same stellar mass distribution. See the values reported in tables \ref{tab:masmetrel_KS_mor} and \ref{tab:masmetdiff_KS_mor}.}
    \label{fig:massmetrel_KS_mor}
\end{figure*}

\begin{figure*}

    \centering
    \includegraphics[width=0.73\linewidth]{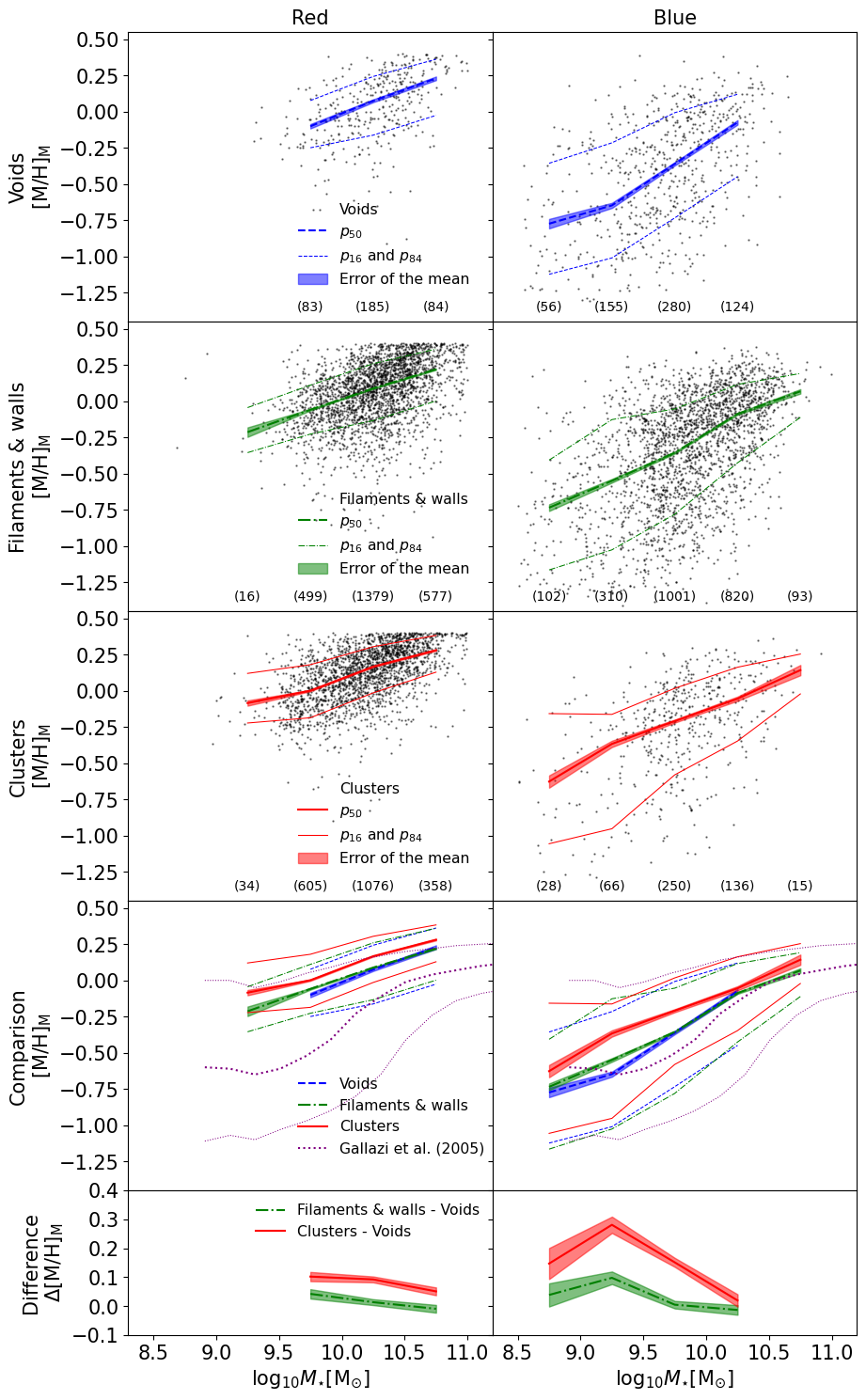}
    \caption{Same as Figure \ref{fig:massmetrel_QC_col}, but for the sub-samples of galaxies with the same stellar mass distribution. See the values reported in tables \ref{tab:masmetrel_KS_col} and \ref{tab:masmetdiff_KS_col}.}
    \label{fig:massmetrel_KS_col}
\end{figure*}

\begin{figure*}

    \centering
    \includegraphics[width=\linewidth]{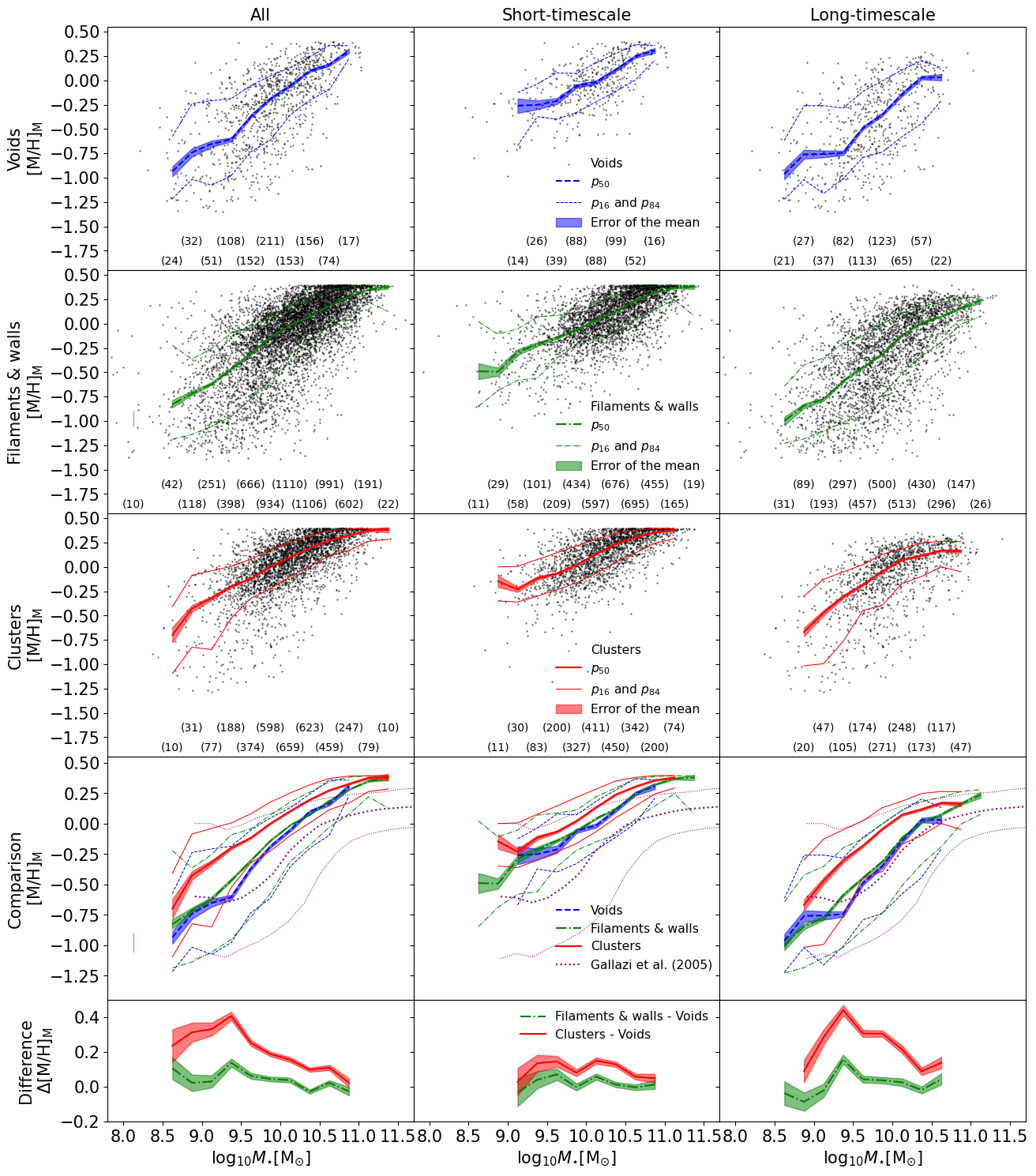}
    \caption{Same as Figure \ref{fig:massmetrel_QC},  but for stellar mass bins of 0.25 dex width. See the values reported in tables \ref{tab:masmetrel_QC_0.25} and \ref{tab:masmetdiff_QC_0.25}.}
    \label{fig:massmetrel_QC_0.25}
\end{figure*}

\begin{figure*}

    \centering
    \includegraphics[width=0.73\linewidth]{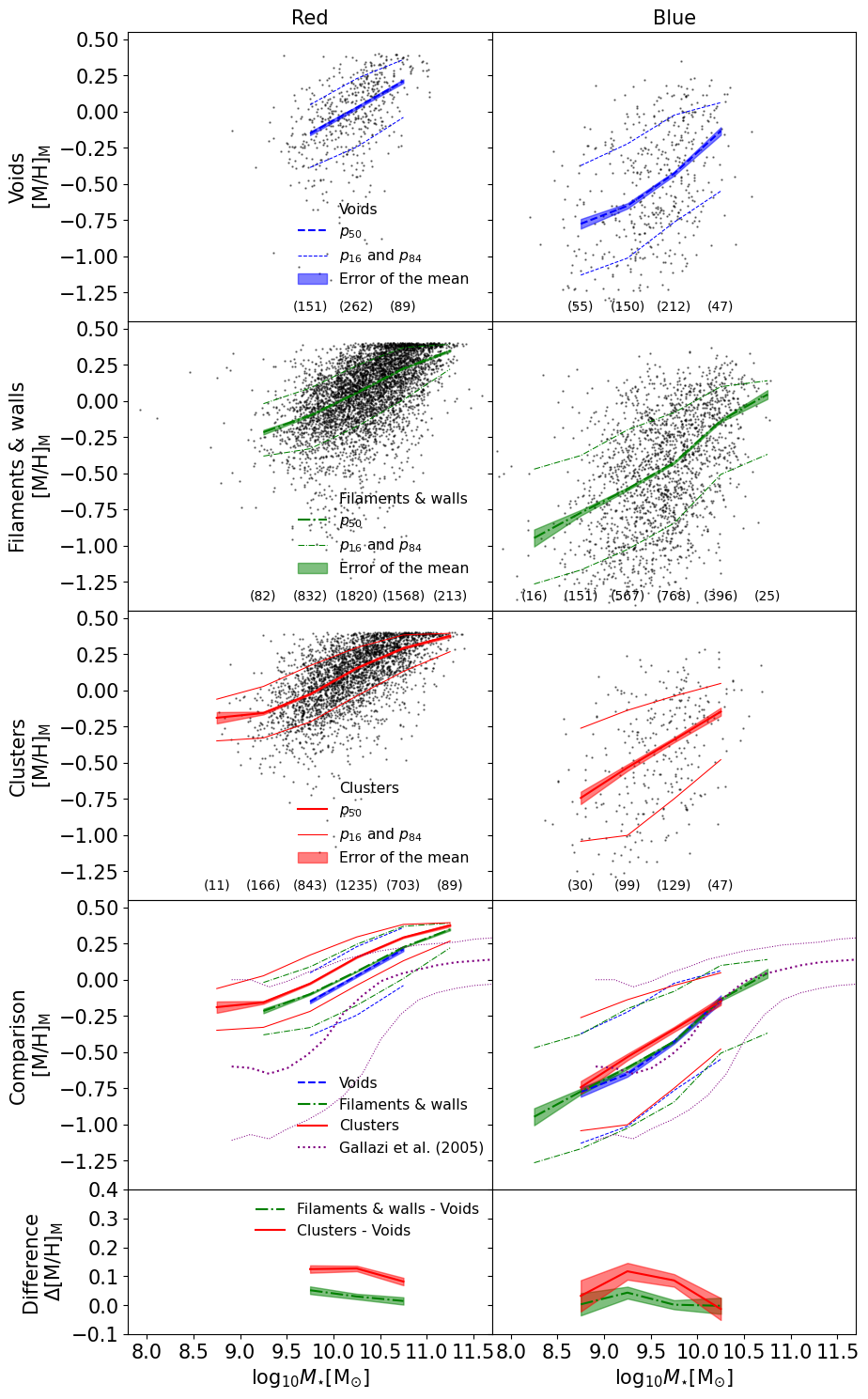}
    \caption{Same as Figure \ref{fig:massmetrel_QC_col}, but with the colour classification criterion at 0.6 instead of 0.7. See the values reported in tables \ref{tab:masmetrel_QC_col0.6} and \ref{tab:masmetdiff_QC_col0.6}.}
    \label{fig:massmetrel_QC_col0.6}
\end{figure*}

In \makebox{figures~\ref{fig:massmetrel_KS}-\ref{fig:massmetrel_KS_col}} we showed the same as \makebox{figures~\ref{fig:massmetrel_QC}-\ref{fig:massmetrel_QC_col},} respectively, but for the sub-samples with the same stellar mass distribution inside each stellar mass bin. The values represented in the figures are reported in \makebox{tables~\ref{tab:masmetrel_KS}-\ref{tab:masmetdiff_KS_col}.} The stellar metallicity differences that we find between galaxies in voids, filaments \& walls, and cluster for the sub-samples with the same stellar mass distribution are similar to what we find in Section~\ref{sec:results} for the main sample. We also show in Figure \ref{fig:massmetrel_QC_0.25} the same as in Figure \ref{fig:massmetrel_QC}, but for narrower stellar mass bins of 0.25 dex width. Additionally, in Figure~\ref{fig:massmetrel_QC_col0.6} we show the same as in Figure~\ref{fig:massmetrel_QC_col} for red and blue galaxies, but with the classification criterion at 0.6 instead of 0.7~mag. The results change, but the general tendency is similar.

\section{Effect of using flux-limited samples from the SDSS}\label{sec:flux-limited}

By definition, SDSS is a flux-limited sample, that is, only galaxies with $r < 17.77~{\rm mag}$ are observed. This magnitude limit prevents the observation of faint galaxies at large distances \citep[see Fig. 2 of][]{2014A&A...566A...1T}, and together with the cluster definition adopted in this paper, this might induce a misclassification of galaxies belonging to clusters as filaments.

\begin{figure*}[!h]
    \centering
        \includegraphics[width=\linewidth]{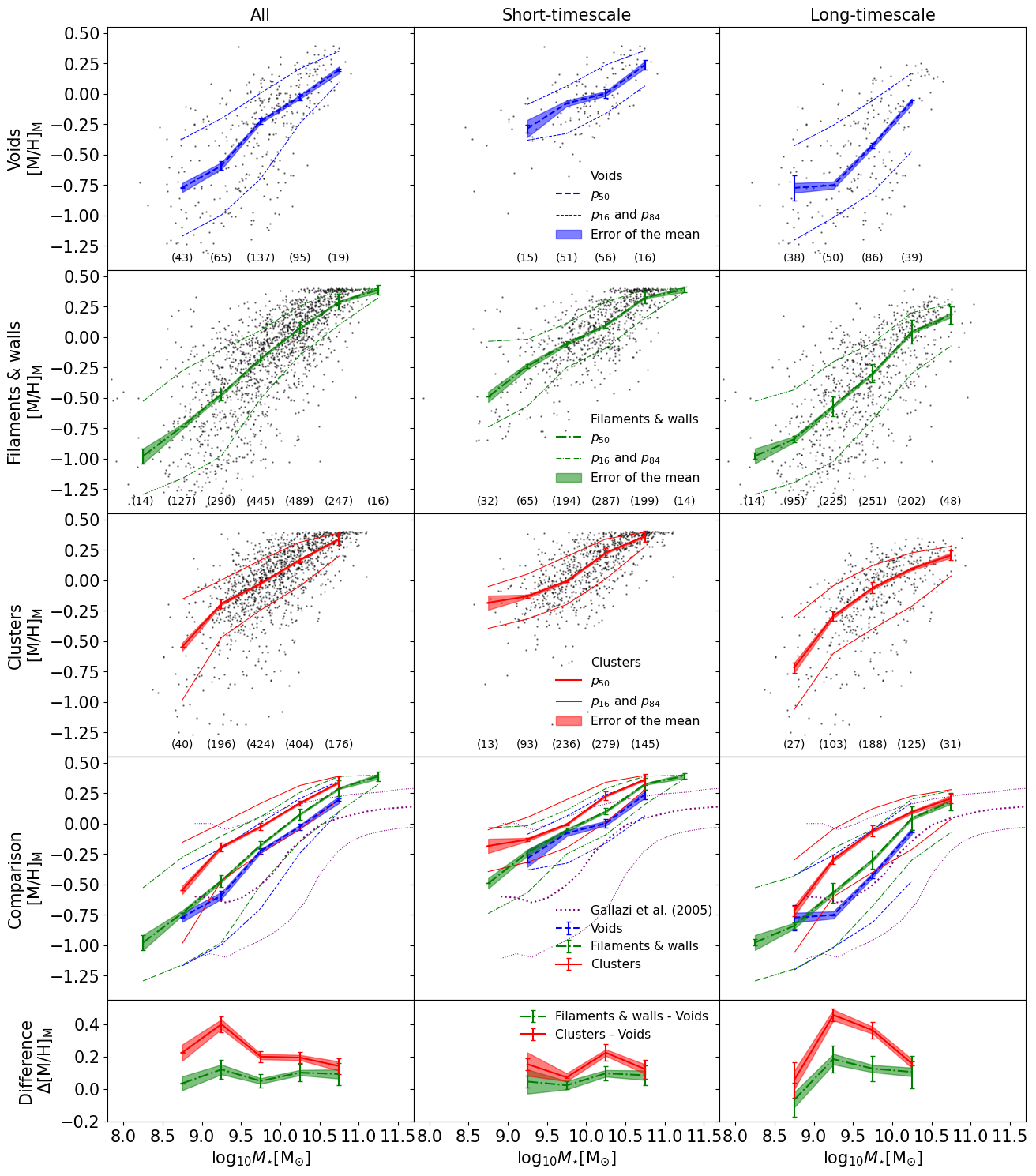}
    \caption{Same as Figure \ref{fig:massmetrel_QC}, but for volume-limited sub-samples of galaxies with redshifts between 0.01 and 0.03. The error bars represent the differences between the results for the flux-limited sub-sample of Figure \ref{fig:massmetrel_QC} and the results for the volume-limited sub-samples of this figure. See the values reported in tables \ref{tab:results_z003_QC} and \ref{tab:results_z003_diff_QC}.}
    \label{fig:results_z003}
    \end{figure*}

In the redshift range analysed in this work, \makebox{$0.01<z<0.05$}, the magnitude limit affects galaxies with absolute magnitude $M_r > -18.25$, which roughly corresponds to stellar masses of $M_\star < 10^{9.5}~{\rm M_\odot}$, overlapping with the stellar mass range of this study, $M_\star > 10^{8.5}~{\rm M_\odot}$. 
In order to test for possible biases introduced by this effect, we analyse in this appendix a volume-limited sub-sample. To do this, we restricted the analysis to redshifts between 0.01 and 0.03, in which case the flux limit of the SDSS produces an uncompleteness of faint galaxies with   $M_r > -17.0$. This magnitude limit roughly corresponds to a limit in stellar mass of  $10^{8.5}~{\rm M_\odot}$, which is the lower end of the masses considered in our analysis and is therefore not expected to be of any relevance for our analysis. The results of our analysis based on this volume-limited sub-sample are shown in Fig.~\ref{fig:results_z003}. Not only the results remain, but the differences are slightly higher, reinforcing the main conclusions of this work. In addition, we show as error bars the differences between the results for the flux-limited sub-samples of Figure \ref{fig:massmetrel_QC} and the results for the volume-limited sub-samples of this figure in order to present an estimate of the uncertainty of our analysis due to possible biases inherited in the sample. The differences reported in this work cannot be accounted for by only considering the systematic errors in the analysis.

This shows that although the use of a flux-limited sample can affect the classification of galaxies as clusters or filaments in principle, in our sample, which was selected from a relatively small redshift range (\makebox{$0.01<z<0.05$}), the effect is small and does not alter the conclusions of our analysis.

\section{Sample selection effect of the cut in signal-to-noise ratio \label{sec:s2n}}

\begin{figure*}

    \centering
    \includegraphics[width=\linewidth]{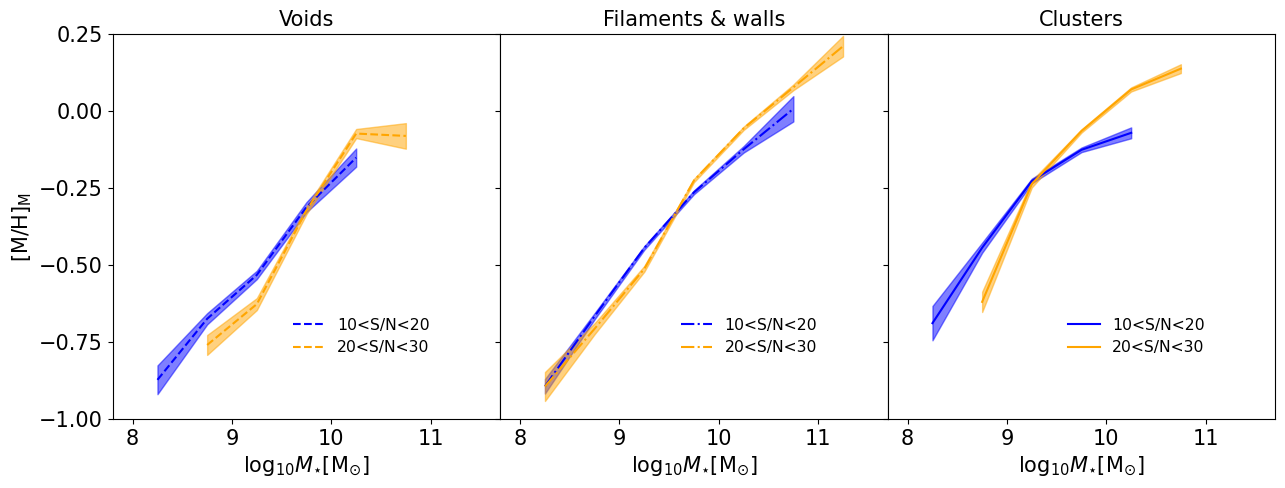}
    \caption{Stellar metallicity vs. stellar mass relation for two ranges of the S/N of the spectra of the galaxies.}
    \label{fig:mass_met_rel_s2n}
\end{figure*}

\begin{figure*}

    \centering
    \includegraphics[width=\linewidth]{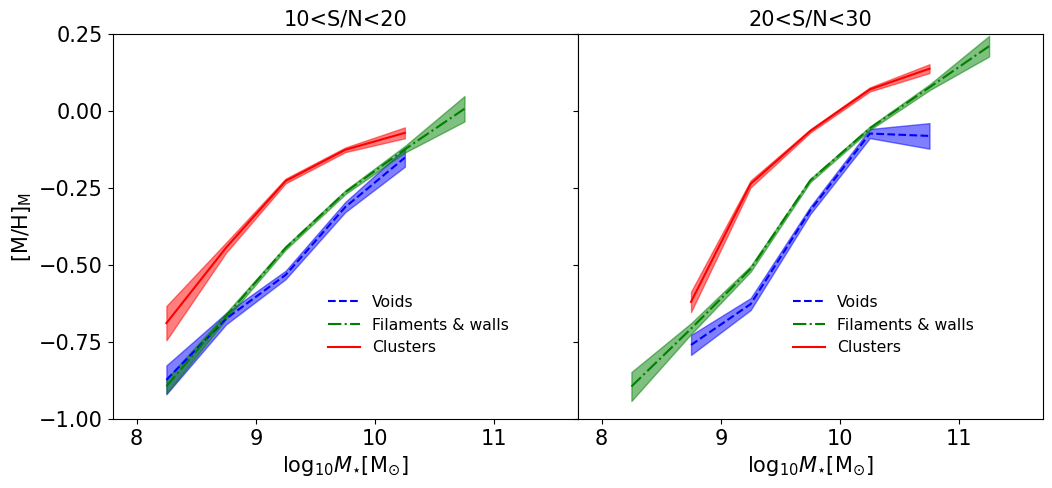}
    \caption{Comparison of the stellar metallicity vs. stellar mass relation between galaxies in voids, filaments \& walls, and clusters for two ranges of the signal-to-noise of the spectra of the galaxies.}
    \label{fig:mass_met_rel_s2n_2}
\end{figure*}

In this section, we test whether the S/N cut of our sample selection might introduce a bias in  our results. We show in Figure \ref{fig:mass_met_rel_s2n} the relation of stellar metallicity versus stellar mass for two ranges of the S/N of the spectra of the galaxies. In Figure \ref{fig:mass_met_rel_s2n_2}, we compare the stellar mass-stellar metallicity relation between galaxies in voids, filaments \& walls, and clusters for the same two ranges of the S/N. The slope of the stellar mass-metallicity relation changes little with the S/N for galaxies in voids and filaments \& walls, but the change is more significant for galaxies in clusters. However, Figure \ref{fig:mass_met_rel_s2n_2} shows that this change does not affect our main result, that is, we find a significant difference in stellar metallicity between galaxies in different large-scale environments for both ranges of the S/N.

\section{Luminosity-weighted stellar mass-metallicity relation\label{sec:MZLR}}

\begin{figure*}

    \centering
    \includegraphics[width=\linewidth]{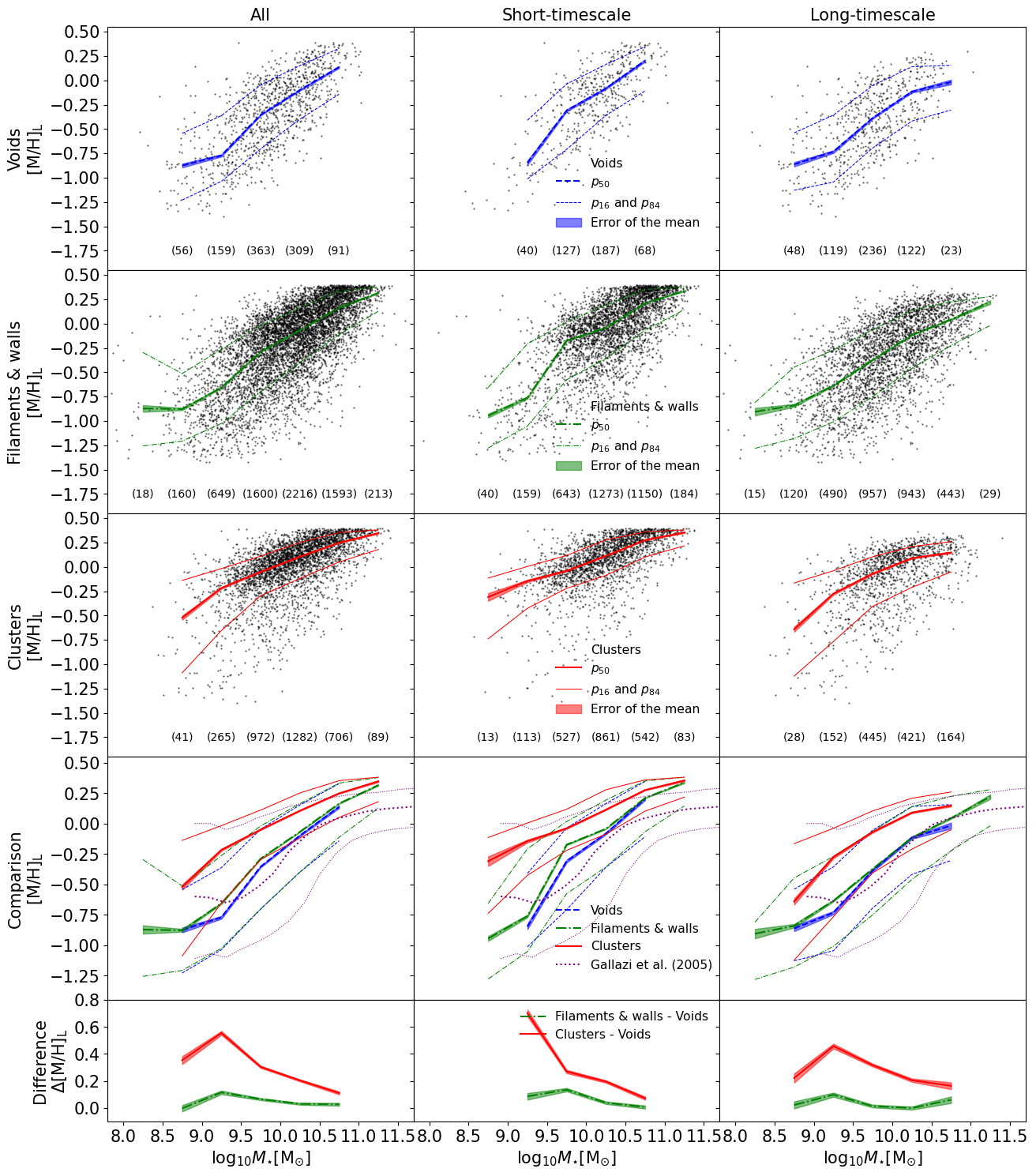}
    \caption{Same as Figure \ref{fig:massmetrel_QC}, but for the stellar metallicity calculated as the luminosity-weighted average. See the values reported in tables \ref{tab:masmetrel_L} and \ref{tab:masmetdiff_L}.}
    \label{fig:massmetLrel_QC}
\end{figure*}

\begin{figure*}

    \centering
    \includegraphics[width=0.73\linewidth]{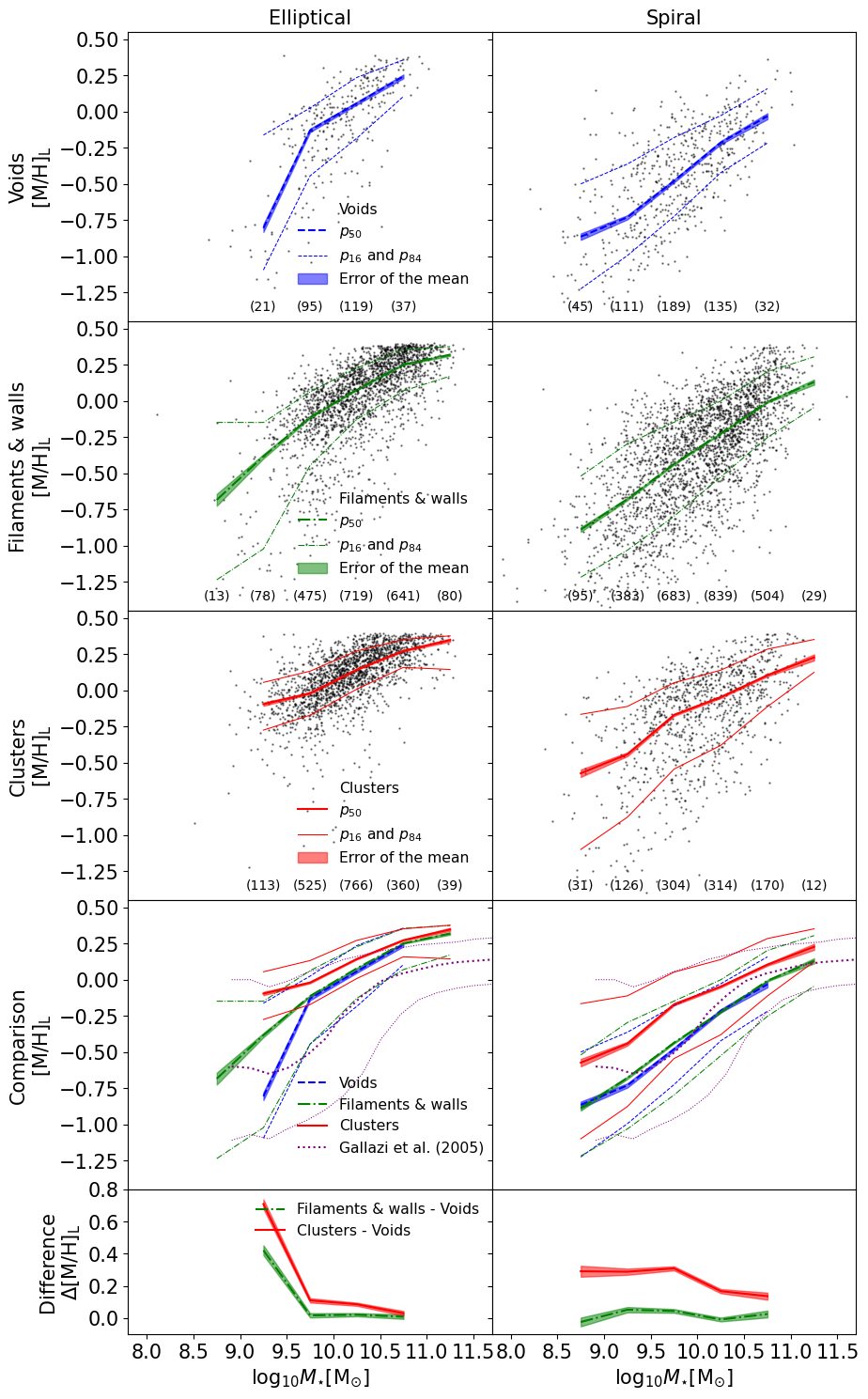}
    \caption{Same as Figure \ref{fig:massmetrel_QC_mor}, but for the stellar metallicity calculated as the luminosity-weighted average. See the values reported in tables \ref{tab:masmetrel_L_mor} and \ref{tab:masmetdiff_L_mor}.}
    \label{fig:massmetLrel_QC_mor}
\end{figure*}

\begin{figure*}

    \centering
    \includegraphics[width=0.73\linewidth]{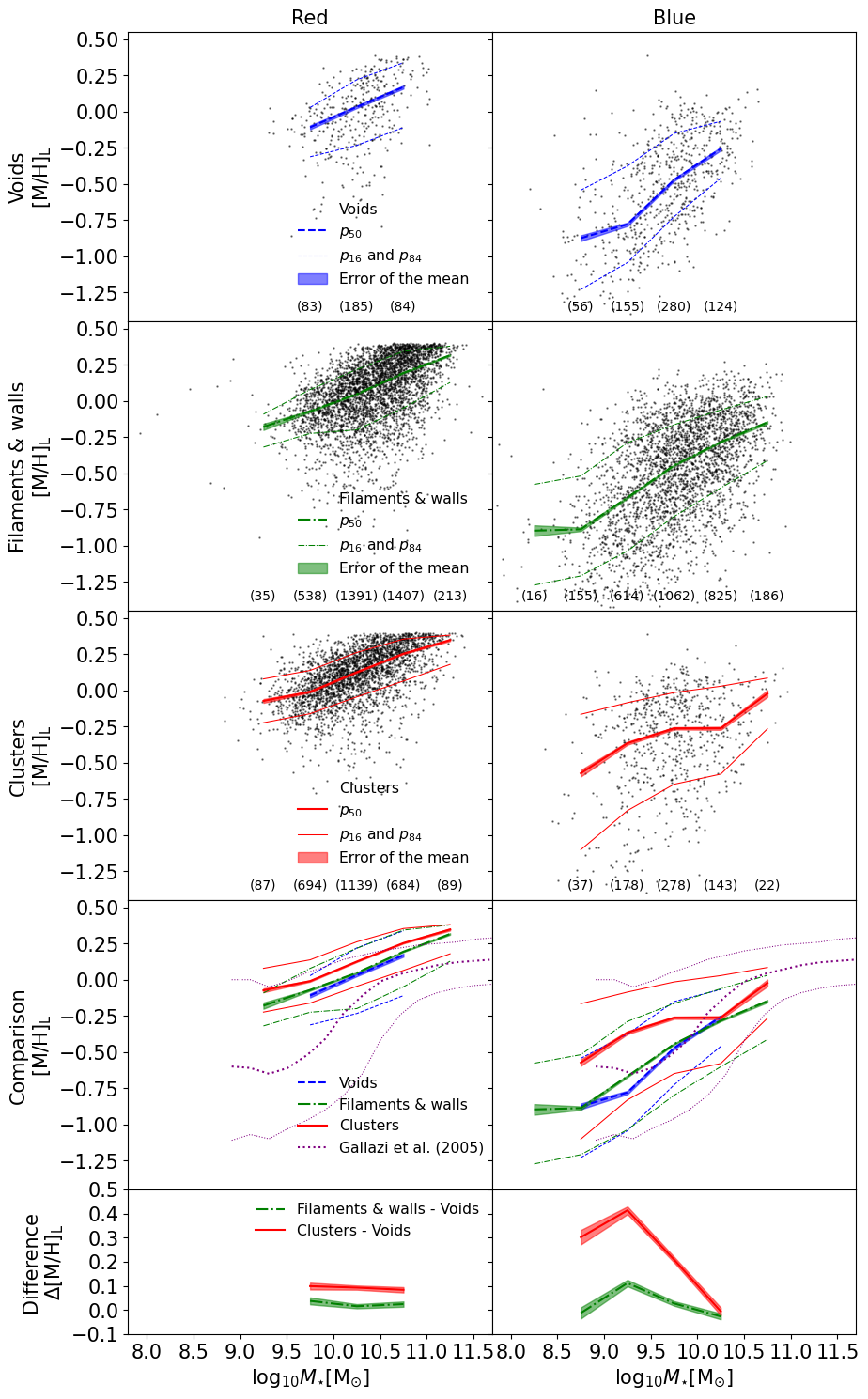}
    \caption{Same as Figure \ref{fig:massmetrel_QC_col}, but for the stellar metallicity calculated as the luminosity-weighted average. See the values reported in tables \ref{tab:masmetrel_L_col} and \ref{tab:masmetdiff_L_col}.}
    \label{fig:massmetLrel_QC_col}
\end{figure*}

In this section, we present the $\rm MZ_\star R$ for different SFH types (Figure~\ref{fig:massmetLrel_QC}), morphologies (Figure~\ref{fig:massmetLrel_QC_mor}), and colours (Figure~\ref{fig:massmetLrel_QC_col}) when the stellar metallicity of a galaxy is calculated as the luminosity-weighted average,

$$\rm [M/H]_L=\frac{\sum L_\star [M/H]_\star}{\sum L_\star},$$

\noindent where $\rm L_\star$ and $\rm [M/H]_\star$ are the luminosity and metallicity of the stars within the galaxy.

The luminosity-weighted stellar metallicities differences that we find between galaxies in different large-scale environments are similar to what we find in Section \ref{sec:results} for the mass-weighted average.

\section{Tables\label{sec:tables}}
In this section, we report the $\rm MZ_\star R$ in tables. In tables~\ref{tab:masmetrel_QC} and \ref{tab:masmetdiff_QC}, we present the values represented in Figure~\ref{fig:massmetrel_QC} for all the galaxies, galaxies with ST-SFH, and galaxies with LT-SFH. In tables~ \ref{tab:masmetrel_QC_mor} and \ref{tab:masmetdiff_QC_mor}, we present the values represented in Figure~\ref{fig:massmetrel_QC_mor} for elliptical and spiral galaxies. In tables~\ref{tab:masmetrel_QC_col} and \ref{tab:masmetdiff_QC_col}, we present the values represented in Figure~\ref{fig:massmetrel_QC_col} for red and blue galaxies with the classification criterion at 0.7~mag. In \makebox{tables~\ref{tab:masmetrel_KS}-\ref{tab:masmetdiff_KS_col}}, we present the values represented in \makebox{figures~\ref{fig:massmetrel_KS}-\ref{fig:massmetrel_KS_col}} for the sub-samples with the same stellar mass distribution. In tables \ref{tab:masmetrel_QC_0.25} and \ref{tab:masmetdiff_QC_0.25}, we present the values represented in Figure~\ref{fig:massmetrel_QC_0.25} for narrower stellar mass bins with a width of 0.25 dex. In tables~\ref{tab:masmetrel_QC_col0.6} and \ref{tab:masmetdiff_QC_col0.6}, we present the values represented in Figure~\ref{fig:massmetrel_QC_col0.6} for red and blue galaxies with the classification criterion at 0.6~mag. \makebox{In tables~\ref{tab:masmetrel_L}-\ref{tab:masmetdiff_L_col}}, we present the values represented in \makebox{figures~\ref{fig:massmetLrel_QC}-\ref{fig:massmetLrel_QC_col}} applying the luminosity-weighted average.

\clearpage

\begin{table*}
    \footnotesize
\caption{\label{tab:masmetrel_QC}Stellar mass-metallicity relation for different SFH types.}
\begin{center}
\begin{tabular}{c|cccc|cccc|cccc}
\multicolumn{13}{c}{$\rm [M/H]_M$}\\
\hline
& \multicolumn{4}{c|}{All}& \multicolumn{4}{c|}{ST-SFH}& \multicolumn{4}{c}{LT-SFH}\\
      $ \log_{10}M_\star[{\rm M_\odot}]$ & n & $p_{50}$ & $p_{16}$ & $p_{84}$ & n &  $p_{50}$ & $p_{16}$ & $p_{84}$ & n & $p_{50}$ & $p_{16}$ & $p_{84}$\\
      \hline\hline
      \multicolumn{13}{l}{(a) Voids}\\
      \hline
       8.25 &             4 &               - &               - &               - &                         1 &                           - &                           - &                           - &                        3 &                          - &                          - &                          - \\
       8.75 &            56 &    -0.77$\pm$0.03 &           -1.12 &           -0.36 &                         8 &                           - &                           - &                           - &                       48 &               -0.88$\pm$0.04 &                      -1.18 &                      -0.48 \\
       9.25 &           159 &  -0.627$\pm$0.017 &           -1.01 &           -0.20 &                        40 &                -0.25$\pm$0.04 &                       -0.42 &                       -0.07 &                      119 &             -0.752$\pm$0.019 &                      -1.06 &                      -0.28 \\
       9.75 &           363 &  -0.247$\pm$0.009 &           -0.70 &            0.01 &                       127 &              -0.076$\pm$0.016 &                       -0.34 &                        0.08 &                      236 &             -0.403$\pm$0.012 &                      -0.77 &                      -0.05 \\
      10.25 &           309 &  -0.000$\pm$0.009 &           -0.29 &            0.21 &                       187 &               0.038$\pm$0.011 &                       -0.16 &                        0.24 &                      122 &             -0.072$\pm$0.014 &                      -0.49 &                       0.16 \\
      10.75 &            91 &   0.208$\pm$0.012 &           -0.07 &            0.36 &                        68 &               0.279$\pm$0.013 &                        0.06 &                        0.37 &                       23 &                0.05$\pm$0.03 &                      -0.20 &                       0.14 \\
      11.25 &             4 &               - &               - &               - &                         3 &                           - &                           - &                           - &                        1 &                          - &                          - &                          - \\
      \multicolumn{13}{l}{(b) Filaments \& walls}\\
      \hline
      8.25 &               18 &       -0.92$\pm$0.05 &              -1.24 &              -0.21 &                            3 &                              - &                              - &                              - &                          15 &                  -0.95$\pm$0.06 &                         -1.28 &                         -0.58 \\
       8.75 &              160 &     -0.737$\pm$0.018 &              -1.16 &              -0.31 &                           40 &                   -0.49$\pm$0.04 &                          -0.73 &                          -0.05 &                         120 &                -0.865$\pm$0.021 &                         -1.20 &                         -0.43 \\
       9.25 &              649 &     -0.519$\pm$0.009 &              -1.01 &              -0.13 &                          159 &                 -0.255$\pm$0.017 &                          -0.57 &                           0.05 &                         490 &                 -0.648$\pm$0.01 &                         -1.06 &                         -0.24 \\
       9.75 &             1600 &     -0.211$\pm$0.004 &              -0.66 &               0.04 &                          643 &                 -0.080$\pm$0.006 &                          -0.27 &                           0.11 &                         957 &                -0.374$\pm$0.006 &                         -0.81 &                         -0.05 \\
      10.25 &             2216 &      0.031$\pm$0.003 &              -0.24 &               0.23 &                         1273 &                  0.073$\pm$0.004 &                          -0.12 &                           0.27 &                         943 &                -0.056$\pm$0.005 &                         -0.42 &                          0.15 \\
      10.75 &             1593 &      0.223$\pm$0.003 &               0.00 &               0.37 &                         1150 &                  0.277$\pm$0.003 &                           0.07 &                           0.38 &                         443 &                 0.111$\pm$0.006 &                         -0.13 &                          0.24 \\
      11.25 &              213 &      0.347$\pm$0.007 &               0.22 &               0.39 &                          184 &                  0.370$\pm$0.007 &                           0.24 &                           0.39 &                          29 &                 0.241$\pm$0.022 &                          0.10 &                          0.28 \\
      \multicolumn{13}{l}{(c) Clusters}\\
      \hline
      8.25 &               2 &                 - &                 - &                 - &                           0 &                             - &                             - &                             - &                          2 &                            - &                            - &                            - \\
       8.75 &              41 &      -0.55$\pm$0.03 &             -0.96 &             -0.16 &                          13 &                  -0.18$\pm$0.06 &                         -0.40 &                         -0.05 &                         28 &                 -0.67$\pm$0.04 &                        -1.06 &                        -0.30 \\
       9.25 &             265 &      -0.23$\pm$0.01 &             -0.60 &             -0.01 &                         113 &                -0.146$\pm$0.014 &                         -0.33 &                          0.05 &                        152 &               -0.333$\pm$0.015 &                        -0.85 &                        -0.07 \\
       9.75 &             972 &    -0.054$\pm$0.004 &             -0.27 &              0.16 &                         527 &                -0.012$\pm$0.005 &                         -0.20 &                          0.19 &                        445 &               -0.108$\pm$0.007 &                        -0.42 &                         0.10 \\
      10.25 &            1282 &     0.148$\pm$0.003 &             -0.06 &              0.29 &                         861 &                 0.190$\pm$0.004 &                         -0.02 &                          0.32 &                        421 &                0.089$\pm$0.006 &                        -0.15 &                         0.21 \\
      10.75 &             706 &     0.292$\pm$0.004 &              0.13 &              0.38 &                         542 &                 0.319$\pm$0.004 &                          0.18 &                          0.39 &                        164 &                0.166$\pm$0.009 &                        -0.00 &                         0.26 \\
      11.25 &              89 &     0.376$\pm$0.008 &              0.27 &              0.39 &                          83 &                 0.379$\pm$0.008 &                          0.29 &                          0.39 &                          6 &                            - &                            - &                            - \\
\end{tabular}
\end{center}
\tablefoot{Stellar mass-metallicity relation for all the galaxies regardless of their SFH type (left multi-column), galaxies with ST-SFH (centre multi-column), and galaxies with LT-SFH (right multi-column) in voids (a), filaments \& walls (b), and clusters (c). The stellar mas-metallicity relation is derived as the 50th percentile ($p_{50}$) of the distribution, together with the error of the mean, inside each stellar mass bin. The 16th ($p_{16}$) and 84th ($p_{84}$) percentiles represent the dispersion of the stellar mass-metallicity relation. The number of galaxies (n) inside each stellar mas bin is also reported.}

\end{table*}

\begin{table*}
    \footnotesize
\caption{\label{tab:masmetdiff_QC}Differences of the stellar mass-metallicity relation between galaxies in voids, filaments \& walls, and clusters for different SFH types.}

\begin{center}
\begin{tabular}{c|cc|cc|cc}
\multicolumn{7}{c}{$\rm \Delta [M/H]_M$}\\
\hline
& \multicolumn{2}{c|}{All}& \multicolumn{2}{c|}{ST-SFH} & \multicolumn{2}{c}{LT-SFH}\\
      $ \log_{10}M_\star[{\rm M_\odot}]$  & $\Delta p_{50}$ & $\sigma$ & $\Delta p_{50}$ & $\sigma$ & $\Delta p_{50}$ & $\sigma$\\
      \hline\hline
      \multicolumn{7}{l}{(a) Filaments \& walls - Voids}\\
      \hline
       8.25 &                          - &                                - &                                      - &                                            - &                                     - &                                           - \\
       8.75 &                0.04$\pm$0.04 &                              1.0 &                                      - &                                            - &                           0.01$\pm$0.04 &                                         0.2 \\
       9.25 &              0.108$\pm$0.019 &                              5.7 &                           -0.00$\pm$0.04 &                                         -0.0 &                         0.104$\pm$0.021 &                                         5.0 \\
       9.75 &                0.04$\pm$0.01 &                              4.0 &                         -0.004$\pm$0.017 &                                         -0.2 &                         0.029$\pm$0.013 &                                         2.2 \\
      10.25 &              0.031$\pm$0.009 &                              3.4 &                          0.035$\pm$0.011 &                                          3.2 &                         0.016$\pm$0.015 &                                         1.1 \\
      10.75 &              0.015$\pm$0.013 &                              1.2 &                         -0.003$\pm$0.014 &                                         -0.2 &                           0.06$\pm$0.03 &                                         2.0 \\
      11.25 &                          - &                                - &                                      - &                                            - &                                     - &                                           - \\
      \multicolumn{7}{l}{(b) Clusters - Voids}\\
      \hline
       8.25 &                         - &                               - &                                     - &                                           - &                                    - &                                          - \\
       8.75 &               0.23$\pm$0.05 &                             4.6 &                                     - &                                           - &                          0.20$\pm$0.05 &                                        4.0 \\
       9.25 &              0.40$\pm$0.02 &                            20.0 &                           0.11$\pm$0.04 &                                         2.8 &                        0.419$\pm$0.024 &                                       17.5 \\
       9.75 &               0.19$\pm$0.01 &                            19.0 &                         0.063$\pm$0.017 &                                         3.7 &                        0.295$\pm$0.014 &                                       21.1 \\
      10.25 &             0.148$\pm$0.009 &                            16.4 &                         0.152$\pm$0.011 &                                        13.8 &                        0.161$\pm$0.015 &                                       10.7 \\
      10.75 &             0.084$\pm$0.013 &                             6.5 &                         0.040$\pm$0.014 &                                         2.9 &                          0.11$\pm$0.03 &                                        3.7 \\
      11.25 &                         - &                               - &                                     - &                                           - &                                    - &                                          - \\
      
\end{tabular}
\end{center}
\tablefoot{Differences of stellar mass-metallicity relation between galaxies in filaments \& walls and voids (a), and between galaxies in clusters and voids (b) for all the galaxies, regardless of their SFH type (left multi-column), galaxies with ST-SFH (centre multi-column), and galaxies with LT-SFH (right multi-column). $\Delta p_{50}$ represents the difference (together with its error) of the 50th percentile of the stellar mass-metallicity distribution between different large-scale environments, and $\sigma$ is the ratio of the nominal value and the error or the difference.}

\end{table*}

\begin{table*}
    \footnotesize
\caption{\label{tab:masmetrel_QC_mor}Stellar mass-metallicity relation for different morphologies.}
\begin{center}
\begin{tabular}{c|cccc|cccc}
\multicolumn{9}{c}{$\rm [M/H]_M$}\\
\hline
& \multicolumn{4}{c|}{Elliptical}& \multicolumn{4}{c}{Spiral}\\
      $ \log_{10}M_\star[{\rm M_\odot}]$ & n &  $p_{50}$ & $p_{16}$ & $p_{84}$ & n & $p_{50}$ & $p_{16}$ & $p_{84}$\\
      \hline\hline
      \multicolumn{9}{l}{(a) Voids}\\
      \hline
       8.25 &                    0 &                      - &                      - &                      - &                3 &                  - &                  - &                  - \\
       8.75 &                    3 &                      - &                      - &                      - &               45 &       -0.72$\pm$0.04 &              -1.10 &              -0.26 \\
       9.25 &                   21 &           -0.41$\pm$0.04 &                  -1.02 &                  -0.19 &              111 &     -0.610$\pm$0.022 &              -1.01 &              -0.21 \\
       9.75 &                   95 &         -0.151$\pm$0.015 &                  -0.35 &                   0.04 &              189 &     -0.399$\pm$0.014 &              -0.75 &               0.01 \\
\end{tabular}
\end{center}
\tablefoot{Same as Table \ref{tab:masmetrel_QC}, but for elliptical (left multi-column) and spiral galaxies (right multi-column).}

\end{table*}

\begin{table*}
    \footnotesize
\caption{\label{tab:masmetdiff_QC_mor}Differences of the stellar mass-metallicity relation between galaxies in voids, filaments \& walls, and clusters for different morphologies.}
\begin{center}
\begin{tabular}{c|cc|cc}
\multicolumn{5}{c}{$\rm \Delta [M/H]_M$}\\
\hline
& \multicolumn{2}{c|}{Elliptical} & \multicolumn{2}{c}{Spiral}\\
      $ \log_{10}M_\star[{\rm M_\odot}]$  & $\Delta p_{50}$ & $\sigma$ & $\Delta p_{50}$ & $\sigma$\\
      \hline\hline
      \multicolumn{5}{l}{(a) Filaments \& walls - Voids}\\
      \hline
       8.25 &                                 - &                                       - &                             - &                                   - \\
       8.75 &                                 - &                                       - &                  -0.11$\pm$0.04 &                                -2.8 \\
       9.25 &                       0.05$\pm$0.05 &                                     1.0 &                 0.039$\pm$0.024 &                                 1.6 \\
       9.75 &                     0.057$\pm$0.017 &                                     3.4 &                 0.071$\pm$0.016 &                                 4.4 \\
\end{tabular}
\end{center}
\tablefoot{Same as Table \ref{tab:masmetdiff_QC}, but for elliptical (left multi-column) and spiral galaxies (right multi-column).}

\end{table*}

\begin{table*}
    \footnotesize
\caption{\label{tab:masmetrel_QC_col}Stellar mass-metallicity relation for different colours.}
\begin{center}
\begin{tabular}{c|cccc|cccc}
\multicolumn{9}{c}{$\rm [M/H]_M$}\\
\hline
& \multicolumn{4}{c|}{Red ($g-r>0.7$)}& \multicolumn{4}{c}{Blue ($g-r\leq0.7$)}\\
      $ \log_{10}M_\star[{\rm M_\odot}]$ & n &  $p_{50}$ & $p_{16}$ & $p_{84}$ & n & $p_{50}$ & $p_{16}$ & $p_{84}$\\
      \hline\hline
      \multicolumn{9}{l}{(a) Voids}\\
      \hline
       8.25 &             0 &               - &               - &               - &              4 &                - &                - &                - \\
       8.75 &             0 &               - &               - &               - &             56 &     -0.77$\pm$0.03 &            -1.12 &            -0.36 \\
       9.25 &             4 &               - &               - &               - &            155 &   -0.648$\pm$0.018 &            -1.01 &            -0.22 \\
       9.75 &            83 &  -0.101$\pm$0.015 &           -0.25 &            0.08 &            280 &   -0.361$\pm$0.011 &            -0.73 &            -0.01 \\
\end{tabular}
\end{center}
\tablefoot{Same as Table \ref{tab:masmetrel_QC}, but for red (left multi-column) and blue galaxies (right multi-column).}

\end{table*}

\begin{table*}
    \footnotesize
\caption{\label{tab:masmetdiff_QC_col}Differences of the stellar mass-metallicity relation between galaxies in voids, filaments \& walls, and clusters for different colours.}
\begin{center}
\begin{tabular}{c|cc|cc}
\multicolumn{5}{c}{$\rm \Delta [M/H]_M$}\\
\hline
& \multicolumn{2}{c|}{Red ($g-r>0.7$)} & \multicolumn{2}{c}{Blue ($g-r\leq0.7$)}\\
      $ \log_{10}M_\star[{\rm M_\odot}]$  & $\Delta p_{50}$ & $\sigma$ & $\Delta p_{50}$ & $\sigma$\\
      \hline\hline
      \multicolumn{5}{l}{(a) Filaments \& walls - Voids}\\
      \hline
      8.25 &                          - &                                - &                           - &                                 - \\
       8.75 &                          - &                                - &                 0.03$\pm$0.04 &                               0.8 \\
       9.25 &                          - &                                - &                0.091$\pm$0.020 &                               4.5 \\
       9.75 &              0.044$\pm$0.016 &                              2.8 &               0.010$\pm$0.013 &                               0.8 \\
\end{tabular}
\end{center}
\tablefoot{Same as Table \ref{tab:masmetdiff_QC}, but for red (left multi-column) and blue galaxies (right multi-column).}

\end{table*}

\begin{table*}
    \footnotesize
\caption{\label{tab:masmetrel_KS}Stellar mass-metallicity relation of sub-samples with the same stellar mass distribution for different SFH types.}
\begin{center}
\begin{tabular}{c|cccc|cccc|cccc}
\multicolumn{13}{c}{$\rm [M/H]_M$}\\
\hline
& \multicolumn{4}{c|}{All}& \multicolumn{4}{c|}{ST-SFH}& \multicolumn{4}{c}{LT-SFH}\\
      $ \log_{10}M_\star[{\rm M_\odot}]$ & n & $p_{50}$ & $p_{16}$ & $p_{84}$ & n &  $p_{50}$ & $p_{16}$ & $p_{84}$ & n & $p_{50}$ & $p_{16}$ & $p_{84}$\\
      \hline\hline
      \multicolumn{13}{l}{(a) Voids}\\
      \hline
       8.75 &            56 &    -0.77$\pm$0.03 &           -1.12 &           -0.36 &                         8 &                           - &                           - &                           - &                       48 &               -0.88$\pm$0.04 &                      -1.18 &                      -0.48 \\
       9.25 &           159 &  -0.627$\pm$0.017 &           -1.01 &           -0.20 &                        40 &                -0.25$\pm$0.04 &                       -0.42 &                       -0.07 &                      119 &             -0.752$\pm$0.019 &                      -1.06 &                      -0.28 \\
       9.75 &           363 &  -0.247$\pm$0.009 &           -0.70 &            0.01 &                       127 &              -0.076$\pm$0.016 &                       -0.34 &                        0.08 &                      236 &             -0.403$\pm$0.012 &                      -0.77 &                      -0.05 \\
      10.25 &           309 &  -0.000$\pm$0.009 &           -0.29 &            0.21 &                       187 &               0.038$\pm$0.011 &                       -0.16 &                        0.24 &                      122 &             -0.072$\pm$0.014 &                      -0.49 &                       0.16 \\
\end{tabular}
\end{center}
\tablefoot{Same as Table \ref{tab:masmetrel_QC}, but for sub-samples with the same stellar mass distribution.}

\end{table*}

\begin{table*}
    \footnotesize
\caption{\label{tab:masmetdiff_KS}Differences of the stellar mass-metallicity relation of sub-samples with the same stellar mass distribution for different SFH types.}
\begin{center}
\begin{tabular}{c|cc|cc|cc}
\multicolumn{7}{c}{$\rm \Delta [M/H]_M$}\\
\hline
& \multicolumn{2}{c|}{All}& \multicolumn{2}{c}{ST-SFH} & \multicolumn{2}{c}{LT-SFH}\\
      $ \log_{10}M_\star[{\rm M_\odot}]$  & $\Delta p_{50}$ & $\sigma$ & $\Delta p_{50}$ & $\sigma$ & $\Delta p_{50}$ & $\sigma$\\
      \hline\hline
      \multicolumn{7}{l}{(a) Filaments \& walls - Voids}\\
      \hline
       8.75 &                0.04$\pm$0.04 &                              1.0 &                                      - &                                            - &                           0.08$\pm$0.04 &                                         2.0 \\
       9.25 &              0.111$\pm$0.021 &                              5.3 &                            0.00$\pm$0.05 &                                          0.0 &                         0.111$\pm$0.024 &                                         4.6 \\
       9.75 &              0.038$\pm$0.011 &                              3.5 &                         -0.004$\pm$0.017 &                                         -0.2 &                         0.027$\pm$0.013 &                                         2.1 \\
      10.25 &              0.030$\pm$0.009 &                              3.3 &                          0.033$\pm$0.011 &                                          3.0 &                         0.016$\pm$0.015 &                                         1.1 \\
\end{tabular}
\end{center}
\tablefoot{Same as Table \ref{tab:masmetdiff_QC}, but for sub-samples with the same stellar mass distribution.}

\end{table*}

\begin{table*}
    \footnotesize
\caption{\label{tab:masmetrel_KS_mor}Stellar mass-metallicity relation of sub-samples with the same stellar mass distribution for different morphologies.}
\begin{center}
\begin{tabular}{c|cccc|cccc}
\multicolumn{9}{c}{$\rm [M/H]_M$}\\
\hline
& \multicolumn{4}{c|}{Elliptical}& \multicolumn{4}{c}{Spiral}\\
      $ \log_{10}M_\star[{\rm M_\odot}]$ & n &  $p_{50}$ & $p_{16}$ & $p_{84}$ & n & $p_{50}$ & $p_{16}$ & $p_{84}$\\
      \hline\hline
      \multicolumn{9}{l}{(a) Voids}\\
      \hline
       8.75 &                    3 &                      - &                      - &                      - &               45 &       -0.72$\pm$0.04 &              -1.10 &              -0.26 \\
       9.25 &                   21 &           -0.41$\pm$0.04 &                  -1.02 &                  -0.19 &              111 &     -0.610$\pm$0.022 &              -1.01 &              -0.21 \\
       9.75 &                   95 &         -0.151$\pm$0.015 &                  -0.35 &                   0.04 &              189 &     -0.399$\pm$0.014 &              -0.75 &               0.01 \\
      10.25 &                  119 &          0.108$\pm$0.011 &                  -0.12 &                   0.28 &              135 &     -0.048$\pm$0.015 &              -0.31 &               0.12 \\
\end{tabular}
\end{center}
\tablefoot{Same as Table \ref{tab:masmetrel_QC_mor}, but for sub-samples with the same stellar mass distribution.}

\end{table*}

\begin{table*}
    \footnotesize
\caption{\label{tab:masmetdiff_KS_mor}Differences of the stellar mass-metallicity relation of sub-samples with the same stellar mass distribution for different morphologies.}
\begin{center}
\begin{tabular}{c|cc|cc}
\multicolumn{5}{c}{$\rm \Delta [M/H]_M$}\\
\hline
& \multicolumn{2}{c|}{Elliptical} & \multicolumn{2}{c}{Spiral}\\
      $ \log_{10}M_\star[{\rm M_\odot}]$  & $\Delta p_{50}$ & $\sigma$ & $\Delta p_{50}$ & $\sigma$\\
      \hline\hline
      \multicolumn{5}{l}{(a) Filaments \& walls - Voids}\\
      \hline
       8.75 &                                 - &                                       - &                  -0.09$\pm$0.05 &                                -1.8 \\
       9.25 &                       0.05$\pm$0.05 &                                     1.0 &                   0.09$\pm$0.03 &                                 3.0 \\
       9.75 &                     0.058$\pm$0.017 &                                     3.4 &                 0.068$\pm$0.016 &                                 4.2 \\
      10.25 &                     0.001$\pm$0.012 &                                     0.1 &                -0.009$\pm$0.016 &                                -0.6 \\
\end{tabular}
\end{center}
\tablefoot{Same as Table \ref{tab:masmetdiff_QC_mor} but for sub-samples with the same stellar mass distribution.}

\end{table*}

\begin{table*}
    \footnotesize
\caption{\label{tab:masmetrel_KS_col}Stellar mass-metallicity relation of sub-samples with the same stellar mass distribution for different colours.}
\begin{center}
\begin{tabular}{c|cccc|cccc}
\multicolumn{9}{c}{$\rm [M/H]_M$}\\
\hline
& \multicolumn{4}{c|}{Red ($g-r>0.7$)}& \multicolumn{4}{c}{Blue ($g-r\leq0.7$)}\\
      $ \log_{10}M_\star[{\rm M_\odot}]$ & n &  $p_{50}$ & $p_{16}$ & $p_{84}$ & n & $p_{50}$ & $p_{16}$ & $p_{84}$\\
      \hline\hline
      \multicolumn{9}{l}{(a) Voids}\\
      \hline
       8.75 &             0 &               - &               - &               - &             56 &     -0.77$\pm$0.03 &            -1.12 &            -0.36 \\
       9.25 &             4 &               - &               - &               - &            155 &   -0.648$\pm$0.018 &            -1.01 &            -0.22 \\
       9.75 &            83 &  -0.101$\pm$0.015 &           -0.25 &            0.08 &            280 &   -0.361$\pm$0.011 &            -0.73 &            -0.01 \\
      10.25 &           185 &    0.075$\pm$0.01 &           -0.16 &            0.24 &            124 &   -0.074$\pm$0.016 &            -0.45 &             0.12 \\
\end{tabular}
\end{center}
\tablefoot{Same as Table \ref{tab:masmetrel_QC_col}, but for sub-samples with the same stellar mass distribution.}

\end{table*}

\begin{table*}
    \footnotesize
\caption{\label{tab:masmetdiff_KS_col}Differences of the stellar mass-metallicity relation of sub-samples with the same stellar mass distribution for different colours.}
\begin{center}
\begin{tabular}{c|cc|cc}
\multicolumn{5}{c}{$\rm \Delta [M/H]_M$}\\
\hline
& \multicolumn{2}{c|}{Red ($g-r>0.7$)} & \multicolumn{2}{c}{Blue ($g-r\leq0.7$)}\\
      $ \log_{10}M_\star[{\rm M_\odot}]$  & $\Delta p_{50}$ & $\sigma$ & $\Delta p_{50}$ & $\sigma$\\
      \hline\hline
      \multicolumn{5}{l}{(a) Filaments \& walls - Voids}\\
      \hline
      8.75 &                          - &                                - &                 0.04+/-0.04 &                               1.0 \\
       9.25 &                          - &                                - &               0.098+/-0.022 &                               4.5 \\
       9.75 &              0.042+/-0.016 &                              2.6 &               0.005+/-0.013 &                               0.4 \\
      10.25 &                0.01+/-0.01 &                              1.0 &              -0.013+/-0.017 &                              -0.8 \\
      10.75 &             -0.009+/-0.013 &                             -0.7 &                           - &                                 - \\
      \multicolumn{5}{l}{(b) Clusters - Voids}\\
      \hline
       8.75 &                         - &                               - &                0.15+/-0.05 &                              3.0 \\
       9.25 &                         - &                               - &                0.28+/-0.03 &                              9.3 \\
       9.75 &             0.102+/-0.016 &                             6.4 &              0.151+/-0.016 &                              9.4 \\
      10.25 &               0.09+/-0.01 &                             9.0 &              0.020+/-0.021 &                              1.0 \\
      10.75 &             0.051+/-0.014 &                             3.6 &                          - &                                - \\
      
\end{tabular}
\end{center}
\tablefoot{Same as Table \ref{tab:masmetdiff_QC_col}, but for sub-samples with the same stellar mass distribution.}

\end{table*}

\begin{table*}
    \footnotesize
\caption{\label{tab:masmetrel_QC_0.25}Stellar mass-metallicity relation with stellar mass bins of 0.25 dex width for different SFH types.}
\begin{center}
\begin{tabular}{c|cccc|cccc|cccc}
\multicolumn{13}{c}{$\rm [M/H]_M$}\\
\hline
& \multicolumn{4}{c|}{All}& \multicolumn{4}{c|}{ST-SFH}& \multicolumn{4}{c}{LT-SFH}\\
      $ \log_{10}M_\star[{\rm M_\odot}]$ & n & $p_{50}$ & $p_{16}$ & $p_{84}$ & n &  $p_{50}$ & $p_{16}$ & $p_{84}$ & n & $p_{50}$ & $p_{16}$ & $p_{84}$\\
      \hline\hline
      \multicolumn{13}{l}{(a) Voids}\\
      \hline
       8.12 &             1 &               - &               - &               - &                         0 &                           - &                           - &                           - &                        1 &                          - &                          - &                          - \\
       8.38 &             3 &               - &               - &               - &                         1 &                           - &                           - &                           - &                        2 &                          - &                          - &                          - \\
       8.62 &            24 &    -0.94+/-0.05 &           -1.22 &           -0.58 &                         3 &                           - &                           - &                           - &                       21 &               -0.96+/-0.05 &                      -1.22 &                      -0.62 \\
       8.88 &            32 &    -0.74+/-0.04 &           -1.02 &           -0.24 &                         5 &                           - &                           - &                           - &                       27 &               -0.76+/-0.05 &                      -1.02 &                      -0.26 \\
\end{tabular}
\end{center}
\tablefoot{Same as Table \ref{tab:masmetrel_QC}, but for narrower stellar mass bins of 0.25 dex width.}

\end{table*}

\begin{table*}
    \footnotesize
\caption{\label{tab:masmetdiff_QC_0.25}Differences of the stellar mass-metallicity relation between galaxies in voids, filaments \& walls, and clusters with stellar mass bins of 0.25 dex width for different SFH types.}
\begin{center}
\begin{tabular}{c|cc|cc|cc}
\multicolumn{7}{c}{$\rm \Delta [M/H]_M$}\\
\hline
& \multicolumn{2}{c|}{All}& \multicolumn{2}{c|}{ST-SFH} & \multicolumn{2}{c}{LT-SFH}\\
      $ \log_{10}M_\star[{\rm M_\odot}]$  & $\Delta p_{50}$ & $\sigma$ & $\Delta p_{50}$ & $\sigma$ & $\Delta p_{50}$ & $\sigma$\\
      \hline\hline
      \multicolumn{7}{l}{(a) Filaments \& walls - Voids}\\
      \hline
       8.12 &                          - &                                - &                                      - &                                            - &                                     - &                                           - \\
       8.38 &                          - &                                - &                                      - &                                            - &                                     - &                                           - \\
       8.62 &                0.11+/-0.06 &                              1.8 &                                      - &                                            - &                          -0.04+/-0.07 &                                        -0.6 \\
       8.88 &                0.02+/-0.05 &                              0.4 &                                      - &                                            - &                          -0.09+/-0.05 &                                        -1.8 \\
\end{tabular}
\end{center}
\tablefoot{Same as Table \ref{tab:masmetdiff_QC}, but for narrower stellar mass bins of 0.25 dex width.}

\end{table*}

\begin{table*}
    \footnotesize
\caption{\label{tab:masmetrel_QC_col0.6}Stellar mass-metallicity relation for a different classification of the colour of the galaxies.}
\begin{center}
\begin{tabular}{c|cccc|cccc}
\multicolumn{9}{c}{$\rm [M/H]_M$}\\
\hline
& \multicolumn{4}{c|}{Red ($g-r>0.6$)}& \multicolumn{4}{c}{Blue ($g-r\leq0.6$)}\\
      $ \log_{10}M_\star[{\rm M_\odot}]$ & n &  $p_{50}$ & $p_{16}$ & $p_{84}$ & n & $p_{50}$ & $p_{16}$ & $p_{84}$\\
      \hline\hline
      \multicolumn{9}{l}{(a) Voids}\\
      \hline
       8.25 &             0 &               - &               - &               - &              4 &                - &                - &                - \\
       8.75 &             1 &               - &               - &               - &             55 &     -0.78$\pm$0.03 &            -1.13 &            -0.37 \\
       9.25 &             9 &               - &               - &               - &            150 &   -0.652$\pm$0.018 &            -1.01 &            -0.22 \\
       9.75 &           151 &  -0.151$\pm$0.012 &           -0.39 &            0.05 &            212 &   -0.429$\pm$0.014 &            -0.77 &            -0.02 \\
\end{tabular}
\end{center}
\tablefoot{Same as Table \ref{tab:masmetrel_QC_col}, but for the colour classification criterion at 0.6 instead of 0.7.}

\end{table*}

\begin{table*}
    \footnotesize
\caption{\label{tab:masmetdiff_QC_col0.6}Differences of the stellar mass-metallicity relation between galaxies in voids, filaments\& walls, and clusters for a different classification of the colour of the galaxies.}
\begin{center}
\begin{tabular}{c|cc|cc}
\multicolumn{5}{c}{$\rm \Delta [M/H]_M$}\\
\hline
& \multicolumn{2}{c|}{Red ($>0.6$)} & \multicolumn{2}{c}{Blue ($\leq0.6$)}\\
      $ \log_{10}M_\star[{\rm M_\odot}]$  & $\Delta p_{50}$ & $\sigma$ & $\Delta p_{50}$ & $\sigma$\\
      \hline\hline
      \multicolumn{5}{l}{(a) Filaments \& walls - Voids}\\
      \hline
      8.25 &                         - &                               - &                          - &                                - \\
       8.75 &                         - &                               - &                0.03$\pm$0.05 &                              0.6 \\
       9.25 &                         - &                               - &                0.12$\pm$0.03 &                              4.0 \\
       9.75 &             0.125$\pm$0.013 &                             9.6 &              0.086$\pm$0.022 &                              3.9 \\
\end{tabular}
\end{center}
\tablefoot{Same as Table \ref{tab:masmetdiff_QC_col}, but for the colour classification criterion at 0.6 instead of 0.7.}

\end{table*}

\begin{table*}
    \footnotesize
\caption{\label{tab:results_z003_QC}Stellar mass-metallicity relation for volume-limited sub-samples.}
\begin{center}
\begin{tabular}{c|cccc|cccc|cccc}
\multicolumn{13}{c}{$\rm [M/H]_M$}\\
\hline
& \multicolumn{4}{c|}{All}& \multicolumn{4}{c|}{ST-SFH}& \multicolumn{4}{c}{LT-SFH}\\
      $ \log_{10}M_\star[{\rm M_\odot}]$ & n & $p_{50}$ & $p_{16}$ & $p_{84}$ & n &  $p_{50}$ & $p_{16}$ & $p_{84}$ & n & $p_{50}$ & $p_{16}$ & $p_{84}$\\
      \hline\hline
      \multicolumn{13}{l}{(a) Voids}\\
      \hline
       8.25 &             3 &               - &               - &               - &                         1 &                           - &                           - &                           - &                        2 &                          - &                          - &                          - \\
       8.75 &            43 &    -0.77$\pm$0.04 &           -1.17 &           -0.37 &                         5 &                           - &                           - &                           - &                       38 &               -0.77$\pm$0.04 &                      -1.20 &                      -0.43 \\
       9.25 &            65 &    -0.59$\pm$0.03 &           -1.00 &           -0.20 &                        15 &                -0.29$\pm$0.07 &                       -0.38 &                       -0.09 &                       50 &               -0.75$\pm$0.03 &                      -1.02 &                      -0.26 \\
       9.75 &           137 &  -0.226$\pm$0.015 &           -0.70 &            0.01 &                        51 &              -0.079$\pm$0.023 &                       -0.33 &                        0.06 &                       86 &             -0.425$\pm$0.019 &                      -0.81 &                      -0.05 \\
\end{tabular}
\end{center}
\tablefoot{Same as Table \ref{tab:masmetrel_QC}, but for volume-limited sub-samples with redshifts between 0.01 and 0.03.}

\end{table*}

\begin{table*}
    \footnotesize
\caption{\label{tab:results_z003_diff_QC}Differences of the stellar mass-metallicity relation between galaxies in voids, filaments \& walls, and clusters for volume-limited sub-samples.}
\begin{center}
\begin{tabular}{c|cc|cc|cc}
\multicolumn{7}{c}{$\rm \Delta [M/H]_M$}\\
\hline
& \multicolumn{2}{c|}{All}& \multicolumn{2}{c|}{ST-SFH} & \multicolumn{2}{c}{LT-SFH}\\
      $ \log_{10}M_\star[{\rm M_\odot}]$  & $\Delta p_{50}$ & $\sigma$ & $\Delta p_{50}$ & $\sigma$ & $\Delta p_{50}$ & $\sigma$\\
      \hline\hline
      \multicolumn{7}{l}{(a) Filaments \& walls - Voids}\\
      \hline
       8.25 &                          - &                                - &                                      - &                                            - &                                     - &                                           - \\
       8.75 &                0.04$\pm$0.04 &                              1.0 &                                      - &                                            - &                          -0.07$\pm$0.05 &                                        -1.4 \\
       9.25 &                0.12$\pm$0.03 &                              4.0 &                            0.05$\pm$0.07 &                                          0.7 &                           0.18$\pm$0.03 &                                         6.0 \\
       9.75 &              0.050$\pm$0.017 &                              2.9 &                            0.02$\pm$0.03 &                                          0.7 &                         0.126$\pm$0.022 &                                         5.7 \\
\end{tabular}
\end{center}
\tablefoot{Same as Table \ref{tab:masmetdiff_QC}, but for volume-limited sub-samples with redshifts between 0.01 and 0.03.}

\end{table*}

\begin{table*}
    \footnotesize
\caption{\label{tab:masmetrel_L}Luminosity-weighted stellar mass-metallicity relation for different SFH types.}
\begin{center}
\begin{tabular}{c|cccc|cccc|cccc}
\multicolumn{13}{c}{$\rm [M/H]_L$}\\
\hline
& \multicolumn{4}{c|}{All}& \multicolumn{4}{c|}{ST-SFH}& \multicolumn{4}{c}{LT-SFH}\\
      $ \log_{10}M_\star[{\rm M_\odot}]$ & n & $p_{50}$ & $p_{16}$ & $p_{84}$ & n &  $p_{50}$ & $p_{16}$ & $p_{84}$ & n & $p_{50}$ & $p_{16}$ & $p_{84}$\\
      \hline\hline
      \multicolumn{13}{l}{(a) Voids}\\
      \hline
       8.25 &             4 &               - &               - &               - &                         1 &                           - &                           - &                           - &                        3 &                          - &                          - &                          - \\
       8.75 &            56 &   -0.875$\pm$0.02 &           -1.23 &           -0.54 &                         8 &                           - &                           - &                           - &                       48 &             -0.863$\pm$0.022 &                      -1.13 &                      -0.54 \\
       9.25 &           159 &  -0.772$\pm$0.012 &           -1.04 &           -0.36 &                        40 &              -0.847$\pm$0.022 &                       -1.01 &                       -0.41 &                      119 &             -0.733$\pm$0.014 &                      -1.05 &                      -0.36 \\
       9.75 &           363 &  -0.354$\pm$0.007 &           -0.70 &           -0.05 &                       127 &              -0.309$\pm$0.011 &                       -0.71 &                       -0.04 &                      236 &             -0.387$\pm$0.009 &                      -0.69 &                      -0.05 \\
\end{tabular}
\end{center}
\tablefoot{Same as Table \ref{tab:masmetrel_QC}, but for the stellar metallicity calculated as the luminosity-weighted average.}

\end{table*}

\begin{table*}
    \footnotesize
\caption{\label{tab:masmetdiff_L}Differences of the luminosity-weighted stellar mass-metallicity relation between galaxies in voids, filaments \& walls, and clusters for different SFH types.}
\begin{center}
\begin{tabular}{c|cc|cc|cc}
\multicolumn{7}{c}{$\rm \Delta [M/H]_L$}\\
\hline
& \multicolumn{2}{c|}{All}& \multicolumn{2}{c}{ST-SFH} & \multicolumn{2}{c}{LT-SFH}\\
      $ \log_{10}M_\star[{\rm M_\odot}]$  & $\Delta p_{50}$ & $\sigma$ & $\Delta p_{50}$ & $\sigma$ & $\Delta p_{50}$ & $\sigma$\\
      \hline\hline
      \multicolumn{7}{l}{(a) Filaments \& walls - Voids}\\
      \hline
       8.25 &                          - &                                - &                                      - &                                            - &                                     - &                                           - \\
       8.75 &             -0.002$\pm$0.023 &                             -0.1 &                                      - &                                            - &                           0.02$\pm$0.03 &                                         0.7 \\
       9.25 &              0.115$\pm$0.013 &                              8.8 &                          0.086$\pm$0.024 &                                          3.6 &                         0.098$\pm$0.015 &                                         6.5 \\
       9.75 &              0.066$\pm$0.008 &                              8.2 &                          0.135$\pm$0.012 &                                         11.2 &                           0.01$\pm$0.01 &                                         1.0 \\
\end{tabular}
\end{center}
\tablefoot{Same as Table \ref{tab:masmetdiff_QC} but for the stellar metallicity calculated as the luminosity-weighted average.}

\end{table*}

\begin{table*}
    \footnotesize
\caption{\label{tab:masmetrel_L_mor}Luminosity-weighted stellar mass-metallicity relation for different morphologies.}
\begin{center}
\begin{tabular}{c|cccc|cccc}
\multicolumn{9}{c}{$\rm [M/H]_L$}\\
\hline
& \multicolumn{4}{c|}{Elliptical}& \multicolumn{4}{c}{Spiral}\\
      $ \log_{10}M_\star[{\rm M_\odot}]$ & n &  $p_{50}$ & $p_{16}$ & $p_{84}$ & n & $p_{50}$ & $p_{16}$ & $p_{84}$\\
      \hline\hline
      \multicolumn{9}{l}{(a) Voids}\\
      \hline
       8.25 &                    0 &                      - &                      - &                      - &                3 &                  - &                  - &                  - \\
       8.75 &                    3 &                      - &                      - &                      - &               45 &     -0.865$\pm$0.022 &              -1.23 &              -0.50 \\
       9.25 &                   21 &           -0.81$\pm$0.03 &                  -1.10 &                  -0.16 &              111 &     -0.731$\pm$0.015 &              -0.99 &              -0.36 \\
       9.75 &                   95 &         -0.130$\pm$0.012 &                  -0.44 &                   0.02 &              189 &       -0.48$\pm$0.01 &              -0.72 &              -0.18 \\
\end{tabular}
\end{center}
\tablefoot{Same as Table \ref{tab:masmetrel_QC_mor}, but for the stellar metallicity calculated as the luminosity-weighted average.}
\label{tab:masmetrel_L_mor}
\end{table*}

\begin{table*}
    \footnotesize
\caption{\label{tab:masmetdiff_L_mor}Differences of the luminosity-weighted stellar mass-metallicity relation between galaxies in voids, filaments \& walls, and clusters for different morphologies.}
\begin{center}
\begin{tabular}{c|cc|cc}
\multicolumn{5}{c}{$\rm \Delta [M/H]_L$}\\
\hline
& \multicolumn{2}{c|}{Elliptical} & \multicolumn{2}{c}{Spiral}\\
      $ \log_{10}M_\star[{\rm M_\odot}]$  & $\Delta p_{50}$ & $\sigma$ & $\Delta p_{50}$ & $\sigma$\\
      \hline\hline
      \multicolumn{5}{l}{(a) Filaments \& walls - Voids}\\
      \hline
       8.25 &                                 - &                                       - &                             - &                                   - \\
       8.75 &                                 - &                                       - &                  -0.02$\pm$0.03 &                                -0.7 \\
       9.25 &                       0.42$\pm$0.03 &                                    14.0 &                 0.052$\pm$0.017 &                                 3.1 \\
       9.75 &                     0.017$\pm$0.014 &                                     1.2 &                 0.045$\pm$0.012 &                                 3.8 \\
\end{tabular}
\end{center}
\tablefoot{Same as Table \ref{tab:masmetdiff_QC_mor}, but for the stellar metallicity calculated as the luminosity-weighted average.}

\end{table*}

\begin{table*}
    \footnotesize
\caption{\label{tab:masmetrel_L_col}Luminosity-weighted stellar mass-metallicity relation for different colours.}
\begin{center}
\begin{tabular}{c|cccc|cccc}
\multicolumn{9}{c}{$\rm [M/H]_L$}\\
\hline
& \multicolumn{4}{c|}{Red ($g-r>0.7$)}& \multicolumn{4}{c}{Blue ($g-r\leq0.7$)}\\
      $ \log_{10}M_\star[{\rm M_\odot}]$ & n &  $p_{50}$ & $p_{16}$ & $p_{84}$ & n & $p_{50}$ & $p_{16}$ & $p_{84}$\\
      \hline\hline
      \multicolumn{9}{l}{(a) Voids}\\
      \hline
       8.25 &             0 &               - &               - &               - &              4 &                - &                - &                - \\
       8.75 &             0 &               - &               - &               - &             56 &    -0.875$\pm$0.02 &            -1.23 &            -0.54 \\
       9.25 &             4 &               - &               - &               - &            155 &   -0.779$\pm$0.012 &            -1.04 &            -0.38 \\
       9.75 &            83 &  -0.109$\pm$0.013 &           -0.31 &            0.03 &            280 &   -0.472$\pm$0.008 &            -0.73 &            -0.15 \\
\end{tabular}
\end{center}
\tablefoot{Same as Table \ref{tab:masmetrel_QC_col}, but for the stellar metallicity calculated as the luminosity-weighted average.}

\end{table*}

\begin{table*}
    \footnotesize
\caption{\label{tab:masmetdiff_L_col}Differences of the luminosity-weighted stellar mass-metallicity relation between galaxies in voids, filaments \& walls, and clusters for different colours.}
\begin{center}
\begin{tabular}{c|cc|cc}
\multicolumn{5}{c}{$\rm \Delta [M/H]_L$}\\
\hline
& \multicolumn{2}{c|}{Red ($g-r>0.7$)} & \multicolumn{2}{c}{Blue ($g-r\leq0.7$)}\\
      $ \log_{10}M_\star[{\rm M_\odot}]$  & $\Delta p_{50}$ & $\sigma$ & $\Delta p_{50}$ & $\sigma$\\
      \hline\hline
      \multicolumn{5}{l}{(a) Filaments \& walls - Voids}\\
      \hline
      8.25 &                          - &                                - &                           - &                                 - \\
       8.75 &                          - &                                - &              -0.012$\pm$0.023 &                              -0.5 \\
       9.25 &                          - &                                - &               0.112$\pm$0.013 &                               8.6 \\
       9.75 &              0.038$\pm$0.014 &                              2.7 &               0.028$\pm$0.009 &                               3.1 \\
\end{tabular}
\end{center}
\tablefoot{Same as Table \ref{tab:masmetdiff_QC_col}, but for the stellar metallicity calculated as the luminosity-weighted average.}

\end{table*}

\end{appendix}
%
%

\end{document}